\documentstyle[aps,prb,multicol,epsf]{revtex}
\begin{document}
\title{A diagrammatic approach for open chaotic systems}  
\author{Oded Agam} 
\address{The Racah Institute of Physics, The Hebrew University, 
Jerusalem 91904, Israel.} 
\date{Draft: \today} 
\maketitle 
\begin{abstract} 
A semiclassical diagrammatic approach is constructed for 
calculating correlation 
functions of observables in open chaotic systems with time reversal symmetry. 
The results are expressed in terms of classical correlation functions 
involving Wigner representations of the observables. 
The formalism is used to explain a recent microwave experiment 
on the $4$-disk problem, and
to characterize the two-point function of the photodissociation
cross section of complex molecules. 
\end{abstract} 

\begin{multicols}{2}

\section{Introduction}

An experimental enquiry of the internal structure of a
complex system involves,  usually, some type of a scattering process.
For instance, a photon is scattered from the system and 
then collected by a remote detector. In other situations, 
the collected  objects are fragments of the initial system itself. 
A prototype example of the latter process is the photodissociation 
of molecules: The molecule absorbs a photon, and disintegrate
by redistribution of energy in its vibrational modes\cite{book}.
A common feature of such systems is the coupling to continuum 
modes. Namely, above some energy threshold, the system is open.
 
Open systems are characterized by 
resonances. These are eigenstates of open Hamiltonians 
which are normalizable, and therefore correspond to
complex eigenvalues associated with a decay in time. In most 
interesting situations, this decay is sufficiently slow, 
allowing for the system to explore a large part of the phase space 
before disintegration.  In complex systems, such as nuclei or large 
molecules, the dynamics on these long lived resonances is chaotic,
therefore, the expectation value of a generic observable exhibits
a statistical behavior\cite{Ericsson}. For instance, the absorption 
cross section for photodissociation of large molecules is a pseudorandom 
function of the photon energy\cite{Pack93,Zhang94,Dobbyn95}. This behavior 
suggests a statistical analysis of observables in open chaotic systems
\cite{Porter65,Brody81,Verbaaschot85,Mello85,Smilansky90,Lewenkopf91,Fyodorov97}.

The main purpose of this paper is to construct a diagrammatic 
scheme for calculating correlators of observables in open 
chaotic systems. This diagrammatic approach is similar to the 
cross diagram technic in disordered systems\cite{Abrikosov75}.
However, although both techniques rely on the semiclassical approximation
there are differences: 
The main advantage of the proposed scheme is in its 
capability of describing individual systems rather 
than an ensemble of them.
Ensemble averaging, as opposed to the energy averaging employed here, tends to 
erase ``clean'' features of individual systems. These features may
have important manifestations, as will be demonstrated in this paper.
On the other hand, disorder diagrammatics
provides a systematic way of calculating corrections due to
quantum interference effects. At this stage
we are not able to provide a similar prescription for general chaotic systems. 
Therefore, we confine our attention to open systems with decay 
time shorter than the time at which weak localization effects set in.
The latter time scale, known as the Ehernfest time, diverges logarithmically
in the semiclassical limit\cite{Aleiner97}.  

The semiclassical analysis of a quantum system brings out the relation 
to its underlying classical dynamics\cite{Agam95}. 
For example, the classical dynamics of an electron in a disordered metal 
is diffusive, and ensemble averages of quantum observables of the electron
are expressed in terms of the spectral properties of the diffusion
propagator\cite{Altshuler94}. In more general situations the classical 
evolution is described by the Perron-Frobenius operator whose spectrum,
known as the Ruelle resonances\cite{Ruelle86,Pollicot90}, describes
the irreversible relaxation of probability densities in phase space.

The classical spectrum of the system sets the important time scales
of the problem. When the system is almost close there are two significant
time scales: One is the decay time of the system, $\tau_d$, 
which is inversely proportional to the typical width of the resonances. 
The closer is the system, the longer is the decay time. 
The second time scale, $\tau_c$, is the time which takes for 
a classical density distribution to 
relax to the ergodic state on the energy shell, when the system is closed. 
In diffusive systems this time is known as the Thouless 
time\cite{Thouless74}.  

A large separation between the classical time scales, 
$\tau_d \gg \tau_c$, implies a 
universal statistical behavior of the system on energy scales smaller than
$\hbar/ \tau_c$. Namely, the statistics is described by a 
random matrix theory\cite{Mehta91} suitable for open 
systems\cite{Fyodorov98,Alhassid98}. However,
as will be demonstrated in this paper, there are important manifestations
of the nonuniversal behavior of the system, which are especially
pronounced when $\tau_d$ is of the same 
order as $\tau_c$. The main use of the diagrammatic approach will
be to calculate these individual imprints the system.
This information is most important for
constructing effective models from experimental data.
  
The organization of this paper is as follows. In section 2 we derive the 
classical propagator of the system by energy averaging of
Green functions. This propagator constitutes the basic building block of 
the diagrammatic approach developed in section 3. We shall restrict our
considerations to systems with time reversal symmetry. In section 4
we present two applications of the formalism. One concerns the microwave 
experiment on the $N$-disk scattering system\cite{Lu99}.
The second is the photodissociation 
absorption cross section of complex molecules. In section 5 
we summarize our results, and mention directions of further studies.   

\section{Classical propagation from quantum Green functions}

The purpose of this section is to construct the building block
of the diagrammatic scheme which will be developed in this paper.
It is the classical propagator, which is a generalization of the 
``diffuson'' ladder diagrams of disordered systems\cite{Abrikosov75}. 
Our construction will
rely on the semiclassical approximation for the Green function of
the quantum system\cite{Berry72,Gutzwiller90}. 

\subsection{The semiclassical Green function}

Let $\hat{\cal H}$ be the Hamiltonian  of an open system
having $d$ degrees of freedom, and 
${\cal H}({\bf x})$ be its classical counterpart, where
${\bf x}=({\bf r},{\bf p})$ is a point in the classical phase space.
The advanced ($^+$) and retarded ($^-$) Green functions of the system
are
\begin{eqnarray} 
G^\pm(\epsilon) =\frac{1}{\epsilon\pm i 0 -\hat{\cal H}}, \nonumber
\end{eqnarray}
where $i 0$ denote an infinitesimal positive imaginary part. 
In the semiclassical limit, these Green functions contain two contributions:
\begin{eqnarray} 
G^\pm \simeq  G_W^\pm + G_{osc}^\pm. \nonumber
\end{eqnarray}
The first, known as the Weyl term, 
is a smooth function of the energy given by
\begin{equation} 
\langle {\bf r'} | G_W^\pm(\epsilon)|{\bf r} \rangle  =  
 \int \frac{d{\bf p}}{h^d}
\frac{ e^{\frac{i}{\hbar}{\bf p} \cdot (\bf{r'-r})}} 
{\epsilon \pm i0 -{\cal H}({\bf x})}, \label{Weyl}
\end{equation}
where $h=2 \pi \hbar$ is Planck's constant, and
${\bf x}=(({\bf r}+{\bf r'})/2,{\bf p})$. 
The second contribution is an oscillatory function of the energy expressed 
as a sum over the classical trajectories\cite{Berry72}
from ${\bf r}$ to ${\bf r'}$ with energy $\epsilon$: 
\begin{equation}
\begin{array}{l} 
\langle {\bf r'} | G_{osc}^+(\epsilon) |{\bf r} \rangle =  
\sum_{\mu} A_\mu e^{ \frac{i}{\hbar}  
S_\mu({\bf r'},{\bf r};\epsilon)}, \\
\langle {\bf r} | G_{osc}^-(\epsilon) |{\bf r'} \rangle =  
\sum_{\mu} A_\mu^* e^{ -\frac{i}{\hbar}  
S_\mu({\bf r'},{\bf r};\epsilon)}.
\end{array}                           \label{osc} 
\end{equation}  
Here $S_\mu({\bf r'},{\bf r};\epsilon)$ is the classical action
of the $\mu$-th trajectory, while $A_\mu$ is an amplitude 
which can be expressed as a combination of second derivatives of 
$S_\mu({\bf r'},{\bf r};\epsilon)$ with respect to ${\bf r'}$ and 
${\bf r}$.

\subsection{The classical propagator }
To construct the classical
propagator of the system from an energy average of  Green
functions, consider the following function
\begin{equation}
\Pi({\bf x}',{\bf x};\omega) = \langle \mbox{Tr}\{ G^+(\epsilon \!+\! \hbar
\omega)
\delta ({\bf \hat{x}}\!-\!{\bf x}) G^-(\epsilon) 
\delta ({\bf \hat{x}}\!-\!{\bf x'})\}\rangle, \label{pi0}
\end{equation}
where the $\delta$-function operator is defined by the symmetric
Fourier transform\cite{Berry89}:
\begin{equation}
\delta ({\bf \hat{x}}\!-\!{\bf x}) = 
\int \frac{d {\bf q} d{\bf k}}{h^{2d}}
e^{\frac{i}{\hbar} \left[{\bf k  \cdot}({\bf \hat{r}}-\bf {r})
+ {\bf q  \cdot}({\bf \hat{p}}-\bf {p}) \right]}. \label{deltadefinition}
\end{equation}     
Although not written explicitly, it will be assumed that
these $\delta$-functions have a small finite width, say the integration 
over ${\bf k}$ and ${\bf q}$ is limited to a large hypersphere. 
This width will be taken to zero at the end of the calculation.
Substituting
definition (\ref{deltadefinition}) in (\ref{pi0}) we obtain
\begin{eqnarray}
\Pi({\bf x'},{\bf x};\omega) = \int \frac{ d {\bf q} d {\bf q'}}
{h^{2d}} e^{-\frac{i}{\hbar}({\bf p} \cdot {\bf q} +
{\bf p'} \cdot {\bf q'})} \times~~~~~~~~~~~ \label{pi1} \\
\left\langle
G^+({\bf r'}\!+\!\frac{{\bf q'}}{2},{\bf  r}\!-\! \frac{{\bf q}}{2};
 \epsilon\!+\! \hbar \omega )
G^-({\bf r}\!+\!\frac{{\bf q}}{2}, {\bf r'}\!-\! \frac{{\bf q'}}
{2}; \epsilon )\right\rangle. \nonumber
\end{eqnarray}
Next, we substitute the semiclassical approximation for
the Green functions. Each Green function has 
a smooth and an oscillatory contribution, thus, there are 
four terms in $\Pi({\bf x'},{\bf x};\omega)$. 
However, it is easy to see that two of them, 
$\langle G^\pm_W G^\mp_{osc} \rangle$, vanish upon averaging.
Therefore only the Weyl, $\langle G^+_WG^-_W\rangle$, and 
the oscillatory, $\langle G^+_{osc}G^-_{osc}\rangle$, contributions survive. 

The Weyl term is local in phase space, namely it is significant 
only when ${\bf x} \simeq {\bf x'}$. This term, which we denote
by $\Pi_{loc}({\bf x'},{\bf x};\omega)$, is of minor 
importance for our purposes, and we defer its calculation to Appendix A. 
The result, however, is
\begin{eqnarray}
\Pi_{loc}({\bf x'},{\bf x};\omega) =
\frac{2 \pi }{ -i  \hbar h^d \omega^+} \delta ({\bf x'} -{\bf x})
 \delta(\epsilon - {\cal H}({\bf x}) ), \label{localpi}
\end{eqnarray}
where $\omega^+ =\omega+ i0$.

Consider now the contribution from the oscillatory parts of
the Green functions. The product, $G^+_{osc} G^-_{osc}$, 
consists in a double sum over trajectories: 
$\mu$ from ${\bf  r}\!-\! {\bf q}/2$ to
${\bf r'}\!+\!{\bf q'}/2$ and $\nu$ from  
${\bf r}\!+\!{\bf q}/2$ to  ${\bf r'}\!-\! {\bf q'}/2$, thus
\begin{equation}
G^+_{osc}(\epsilon+\hbar \omega) G^-_{osc}(\epsilon) \simeq 
\sum_{\mu \nu} A_\mu A_{\nu}^* e^{i S_\mu - i S_{\nu}}. \label{doublesum}
\end{equation}
An illustration of two such  orbits is depicted in Fig.~1. 
This term is, clearly, nonlocal in space.  Its oscillatory
dependence on the energy implies that upon energy averaging 
the main contribution comes from 
the diagonal part ($\mu=\nu$) of the double sum (\ref{doublesum}).
The approximation of a double sum
by a single sum is called the ``diagonal approximation''\cite{Berry85}. 
It does not take into account quantum interference corrections such 
as weak localization. However, according to our assumptions on the relation
between the Ehernfest time and the decay time of the system, 
these corrections can be neglected.

{\narrowtext
\begin{figure}[h]
\epsfxsize=8.5cm
%\vspace{-0.5 cm}
\epsfbox{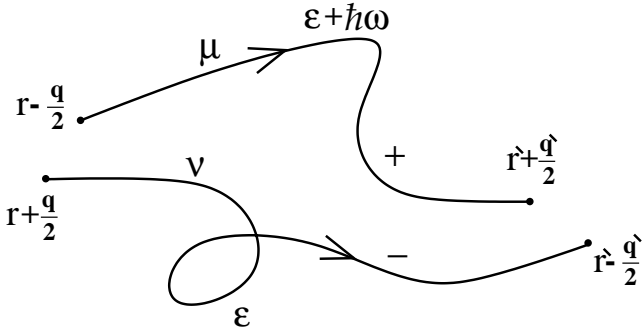}
\vspace{0.5 cm}
\caption{An illustration of the classical trajectories contributing to 
the advanced (+) and the retarded ($-$) Green functions in Eq.~(\ref{pi1}) 
before the energy averaging.}
\label{fig:1}
\end{figure}
}

The construction of the diagonal sum of orbits involves an expansion
of the actions around that of the orbit going from ${\bf r}$ to ${\bf r'}$
namely:
\begin{eqnarray}
S_\mu({\bf r'}\!+\!\frac{{\bf q'}}{2},
{\bf r}\!-\!\frac{{\bf q}}{2};\epsilon+\hbar \omega)& \simeq & 
S_\mu({\bf r'},{\bf r};\epsilon) \nonumber \\
\hbar \omega t_{\mu} &+& \frac{1}{2}\left({\bf p}_{\mu} \cdot
{\bf q}+{\bf p}_{\mu}' \cdot {\bf q'}\right), \nonumber \\ 
S_{\nu}(
{\bf r'}\!-\!\frac{{\bf q'}}{2}, {\bf r}\!+\!\frac{{\bf q}}{2};\epsilon)& 
\simeq& 
S_{\nu}({\bf r'},{\bf r};\epsilon) \nonumber \\  &-& 
\frac{1}{2}\left({\bf p}_{\nu} \cdot
{\bf q}+{\bf p}_{\nu}' \cdot {\bf q'}\right), \nonumber
\end{eqnarray}
where ${\bf p}_{\mu}$ and ${\bf p}_{\mu}'$ denote the initial 
(at point ${\bf r}$)
and the final (at point ${\bf r'}$) momenta associated with the
$\mu$-th orbit. Similarly  ${\bf p}_{\nu}$ and ${\bf p}_{\nu}'$ are 
the initial and the final
momenta associated with the $\nu$-th orbit. $t_{\mu}$ is the time which
takes for the particle to travel along the $\mu$-th trajectory.
Using the above expansions for the actions in (\ref{doublesum}),
the diagonal approximation yields
\begin{eqnarray}
\langle G_{osc}^-(\epsilon+\hbar \omega) G_{osc}^+(\epsilon) \rangle =
\sum_\mu |A_\mu|^2 e^{ \frac{i}{\hbar}\left({\bf p}_{\mu} \cdot {\bf q}
+ {\bf p}_{\mu}' \cdot {\bf q'} + \hbar \omega t_\mu \right)}. \nonumber
\end{eqnarray}
Substituting this result into (\ref{pi1}) and integrating over
${\bf q}$ and ${\bf q'}$ we arrive at
\begin{equation}
\Pi({\bf x'},{\bf x};\omega) =\!\!\!\!\!\!\sum_{\mbox{orbits:}~ {\bf r} 
\to {\bf r'}} \!\!\!\!\!\! |A_\mu|^2 
e^{i\omega t_\mu}\delta({\bf p}\!-\!{\bf p}_{\mu})
\delta({\bf p}'\!-\!{\bf p}_{\mu}'). \label{pi2}
\end{equation} 
This formula expresses $\Pi({\bf x'},{\bf x};\omega)$ as a sum over
the classical trajectories from ${\bf r}$ to ${\bf r'}$. The momenta
$\delta$-functions restrict the initial and final momenta 
of the trajectories to ${\bf  p}$ and ${\bf p'}$, respectively. 

The diagonal sum (\ref{pi2}) can be calculated using the flowing sum 
rule\cite{Agam96} (proved in Appendix B):
\begin{eqnarray} 
h^{d-1}\hbar^2 \!\!\!\!\!\sum_{\mbox{orbits:}~ {\bf r} \to {\bf r'}} \!\!\!\! 
|A_\mu|^2 g({\bf x}_\mu', {\bf x}_\mu, t_\mu) =
~~~~~~~~~~~~~~~~~~~~~ \label{sumrule} \\
\int_0^\infty \! dt d{\bf p'} d{\bf p} g({\bf x'}, {\bf x}, t)  
\delta(\epsilon -{\cal H} ({\bf x})) \delta ({\bf x'}-{\bf x}(t)).
\nonumber 
\end{eqnarray} 
Here $g({\bf x'}, {\bf x}, t)$ is a general function of the
space points ${\bf x'}$ and  ${\bf x}$, and the time $t$.
The coordinates  ${\bf x}_\mu=({\bf r},{\bf p}_\mu)$ and ${\bf x}_\mu'= 
({\bf r}',{\bf p}_\mu')$ are the initial and final phase-space points 
of the $\mu$-th trajectory, while $t_\mu$ is the corresponding period of 
the trajectory. Finally,  ${\bf x}(t)$ is the phase space trajectory
of the system, as function of the time $t$, starting from ${\bf x}$.
This trajectory is the solution of Hamilton's equations:
$\dot{{\bf r}}= \partial{\cal H}({\bf x})/ \partial {\bf p}$,
$\dot{{\bf p}}= -\partial{\cal H}({\bf x})/ \partial {\bf r}$, with initial 
conditions  ${\bf x}(t=0)={\bf x} $. Using (\ref{sumrule}) to replace 
the sum over orbits in (\ref{pi2}) by an integral over the time we finally
obtain 
\begin{equation}
\Pi({\bf x'},{\bf x};\omega) = \delta (\epsilon \!-\!{\cal H}({\bf x}))
\int_0^\infty \!\!\! \frac{dt~ e^{i \omega t} }{h^{d-1}\hbar^2} 
\delta ({\bf x'}\!-\!{\bf x}(t)). \label{pi3}
\end{equation}

The function $\Pi({\bf x'},{\bf x};\omega)$ is the Fourier 
transform of the classical propagator of the system for $t>0$. 
It is the generalization of the ``diffuson'' of disordered systems 
to general chaotic systems. 
This analogy becomes clearer when projecting 
$\Pi({\bf x'},{\bf x};\omega)$ down
to real space, i.e. integrating over ${\bf p}$ and ${\bf p'}$.
 The integration over ${\bf p}'$ is straightforward, and to
integrate over ${\bf p}$ we assume that ${\cal H}({\bf x})= 
p^2/2m$, and that ${\bf r}(t)$ is independent of ${\bf p}$. This is
good approximation for a diffusive motion on long time scales, since
momentum relaxation is fast. Thus
\begin{eqnarray}
\int d {\bf p'} d{\bf p} \Pi({\bf x'},{\bf x};\omega) & \simeq&
\frac{ 2 \pi \bar{\nu}}{\hbar} \int d t e^{ i\omega t} 
\delta ({\bf r'}-{\bf r}(t)), \nonumber
\end{eqnarray}  
where $\bar{\nu}$ is the average density of states per unit volume.
The function, $\delta ({\bf r'}-{\bf r}(t))$,
is the matrix element of the diffusion propagator between the
position states $|{\bf r} \rangle$ and $\langle {\bf r}' |$, i.e.
\begin{eqnarray}
\delta ({\bf r'}-{\bf r}(t)) &=& 
\langle {\bf r'}| e^{D \nabla^2 t}|{\bf r} \rangle, \nonumber
\end{eqnarray}
where $D$ is the diffusion constant. Inserting a complete set 
of momentum states, which 
diagonalizes the propagator, and integrating over $t$ we arrive at formula
for the diffuson of disordered systems:
\begin{eqnarray}
\int d {\bf p'} d{\bf p} \Pi({\bf x'},{\bf x};\omega) & \simeq &
\frac{ 2 \pi \bar{\nu}}{V \hbar}  \sum_q \frac{ e^{i{\bf q}\cdot 
({\bf r'}-{\bf r})}}{ -i \omega 
+ D q^2}, \label{diffusivespectral}
\end{eqnarray}  
where $V$ is the volume of the system. 

A spectral decomposition similar to (\ref{diffusivespectral})
exists also for the classical propagator
of general chaotic systems\cite{Ruelle86,Pollicot90}. 
In the time domain one formally has:
\begin{eqnarray}
\delta({\bf x'}-{\bf x}(t)) &=& \langle {\bf x'} | e^{-{\cal L}t} |{\bf x} 
\rangle \label{decomposition} \\
&=& \delta ({\cal H}({\bf x}) - {\cal H}({\bf x'}))
\sum_\alpha e^{-\gamma_\alpha t} \chi_\alpha^l( {\bf x})
\chi_\alpha^r( {\bf x}'), \nonumber
\end{eqnarray}  
where ${\cal L}=\{~\cdot~, {\cal H}({\bf x})\}$ 
is the Poisson bracket operator of the classical system, 
$\gamma_\alpha$ are its 
eigenvalues, and $\chi_\alpha^l( {\bf x})$ and $\chi_\alpha^r( {\bf x}')$
are the corresponding left and right eigenfunctions. The above spectral
decomposition results from regularization of the classical dynamics. 
For instance, by keeping a finite width for the $\delta$-functions 
in (\ref{pi3}) and taking it to zero only after the diagonalization 
of the propagator. The resulting eigenvalues, called Ruelle resonances,
constitute the Perron-Frobenius spectrum of the classical system.
All classical eigenvalues of open systems 
have positive real part. They appear either as purely real or in complex 
conjugate pairs, We denote this set of eigenvalues by 
$\{\gamma_\alpha \}_{\alpha = 0}^\infty$, and order them 
according to the magnitude of their real parts, $ \gamma_0' \leq  \gamma_1'
 \leq  \gamma_2' \leq \cdots$, where 
$\gamma_\alpha=\gamma_\alpha' \pm i \gamma_\alpha''$.

In concluding this section we remark that complex eigenvalues of the classical
propagator, $\gamma_\alpha= \gamma_\alpha' \pm i \gamma_\alpha''$, 
characterize ballistic chaotic systems. They emerge, 
for example, when typical classical trajectories spend long time near
some short periodic orbit. This type of behavior does not appear in diffusive 
systems where classical relaxation is dominated by purely real eigenvalues. 
 In section 4 we demonstrate the manifestations of
complex Ruelle resonances in two examples. 

\section{Semiclassical diagrammatics for open chaotic systems}

In this section we construct a diagrammatic scheme for calculating 
correlation functions such as the two-point function $\langle \mbox{Tr} 
\{ G^+(\epsilon+ \hbar\omega) \hat{A} \}  \mbox{Tr} 
\{ G^-(\epsilon) \hat{B} \}\rangle$, where $\hat{A}$ and $\hat{B}$ are some 
observables, and the averaging, $\langle \cdots \rangle$, is over 
the ``center of mass'' energy $\epsilon$. The energy interval over which
this averaging takes place is sufficiently wide to  contain a large number
of resonances,
but narrow enough so that the classical dynamics within this 
interval is approximately independent of the energy. 

The proposed diagrammatic scheme is a lift into phase space of the 
disordered diagrammatics which is embedded either in the momentum or real
space. An important step in this direction is to express products of 
quantum observables in terms of their Wigner representations which are
functions of phase space variables. This issue will be considered in
subsection A. Next we calculate the
one-point (subsection B) , and the two-point (subsection C)  functions. 
Finally (subsection D), the results will be generalized to
$n$-point functions by setting the diagram rules for their calculation.

\subsection{Wigner representations}

Let  $\langle {\bf r}|\hat{A}|{\bf r'} \rangle$ be a matrix element
of the observable $\hat{A}$, where $\langle {\bf r}|$ and $|{\bf r'} \rangle$ 
are position states in Dirac notation. 
The Wigner representation\cite{Berry77}, 
\begin{equation}
A({\bf x}) = \int d{\bf q} e^{\frac{i}{\hbar}{\bf p \cdot q}} 
\langle {\bf r}- \frac{\bf q}{2} | \hat{A} |{\bf r}+ \frac{\bf 
q}{2}\rangle, 
\label{wignerQ}
\end{equation}     
of the operator $\hat{A}$ is a function of the  phase space variables 
${\bf x}=({\bf r},{\bf p})$. 
It is a faithful representation of the quantum operator,
since all its matrix elements can be reconstructed from the 
inverse relation
\begin{equation}
\langle {\bf r} | \hat{A} |{\bf r'}\rangle= \int \frac{d{\bf p}}
{h^d}
e^{-\frac{i}{\hbar}{\bf p} \cdot ({\bf r'-r})} 
A({\bf x}),   \label{Iwigner}
\end{equation}
where ${\bf x}= (({\bf r'+r})/2, {\bf p})$.
Yet, expressed as a function on the classical phase space, the Wigner 
representation is convenient for semiclassical expansions.  

The external product of two operators $\hat{A} \otimes \hat{B}$, 
with matrix elements 
$\langle {\bf r}_0 |\hat{A}| {\bf r}_1 \rangle 
\langle {\bf r}_2 |\hat{B}| {\bf r}_3 \rangle$, have more than one 
Wigner representation.
These representations correspond to the various ways by which
pairs of coordinates are used for the Fourier transforms. 
One possibility is to pair the coordinates of each one of
the operators separately, i.e. ${\bf r}_0$ with ${\bf r}_1$ and
${\bf r}_2$ with ${\bf r}_3$. Obviously, the result in this case 
will be the  product of the Wigner representations of $\hat{A}$ and $\hat{B}$. 
Denoting it by $[AB]_d({\bf x},{\bf x'})$, we have
\begin{eqnarray}
[AB]_d({\bf x},{\bf x'}) = A({\bf x})B({\bf x'})=
\int d{\bf q}d{\bf q'}
e^{\frac{i}{\hbar}({\bf p \cdot q+ p'\cdot q' })} \nonumber \\ 
 \times \langle {\bf r}- \frac{\bf q}{2} | 
\hat{A} |{\bf r}+ \frac{\bf q}{2}\rangle \langle {\bf r'}- \frac{\bf q'}{2} 
| \hat{B} |{\bf r'}+ \frac{\bf q'}{2}\rangle,~~~~~~~~~~
\end{eqnarray}
where ${\bf x}= ({\bf r,p})$, and  ${\bf x'}= ({\bf r',p'})$. 
The two other representations correspond to different
pairing configurations: ${\bf r}_0$ with ${\bf r}_3$ and
${\bf r}_1$ with ${\bf r}_2$, which leads to 
\begin{eqnarray}
[AB]_s({\bf x},{\bf x'}) = 
\int d{\bf q}d{\bf q'}
e^{\frac{i}{\hbar}({\bf p \cdot q+ p'\cdot q' })}~~~~~~~~~~~~ \label{ABs}
 \\ 
 \times \langle {\bf r}- \frac{\bf q}{2} | 
\hat{A} |{\bf r'}+ \frac{\bf q'}{2}\rangle \langle {\bf r'}- \frac{\bf q'}{2}
 | \hat{B} |{\bf r}+ \frac{\bf q}{2}\rangle,~~ \nonumber 
\end{eqnarray}
and ${\bf r}_0$ with ${\bf r}_2$,
${\bf r}_1$ with ${\bf r}_3$,  which gives
\begin{eqnarray}
[AB]_c({\bf x},{\bf x'}) = 
\int d{\bf q}d{\bf q'}
e^{\frac{i}{\hbar}({\bf p \cdot q+ p'\cdot q' })}~~~~~~~~~~~~ \label{ABc}
 \\ 
 \times \langle {\bf r}- \frac{\bf q}{2} | 
\hat{A} |{\bf r'}- \frac{\bf 
q'}{2}\rangle \langle {\bf r}+ \frac{\bf q}{2} | \hat{B} 
|{\bf r'}+ \frac{\bf q'}{2}\rangle.~~ \nonumber
\end{eqnarray}

Similar to the inverse relation (\ref{Iwigner}), there are 
inverse relations for products of matrix elements. The simplest case is
the inverse relation for $[AB]_d({\bf x},{\bf x'})=A({\bf x})B({\bf x'})$. 
It is a product of two inverse formulae as (\ref{Iwigner}). The inverse 
relations corresponding to (\ref{ABs}) and (\ref{ABc}) are:
\begin{eqnarray}
\langle {\bf r}_0 |\hat{A}| {\bf r}_1 \rangle 
\langle {\bf r}_2 |\hat{B}| {\bf r}_3 \rangle = 
~~~~~~~~~~~~~~~~~~~~~~~~~~~~~~~~~~~~ \nonumber \\
~~~= \int \frac{d{\bf p}d{\bf p'}}{h^{2d}}
[AB]_s({\bf x},{\bf x'}) e^{-\frac{i}{\hbar}[{\bf p} \cdot 
({\bf r}_3-{\bf r}_0) + {\bf p'} \cdot (\bf{ r}_1 -{\bf r_2})] } 
\label{Wigners}\\
~~~= \int \frac{d{\bf p}d{\bf p'}}{h^{2d}}
[AB]_c({\bf x},{\bf x'}) e^{-\frac{i}{\hbar}[{\bf p} \cdot 
({\bf r}_2-{\bf r}_0) + {\bf p'} \cdot (\bf{ r}_3 -{\bf r_1})] },
\label{Wignerc}
\end{eqnarray}
where in (\ref{Wigners}) ${\bf x}\!=
\!(({\bf r}_3\!+\!{\bf r}_0)/2,{\bf p})$ 
and ${\bf x}'\!=\!(({\bf r}_1\!+\!{\bf r}_2)/2,{\bf p}')$, 
while in (\ref{Wignerc})
${\bf x}\!=\!(({\bf r}_2\!+\!{\bf r}_0)/2,{\bf p})$, and 
${\bf x}'\!=\!(({\bf r}_1\!+\!{\bf r}_3)/2,{\bf p}')$. 

In understanding the nature of the new Wigner representations, it
will be instructive to consider examples. As a first example
consider the operators
\begin{equation}
\hat{A}= f_\sigma (\hat{\bf r}-{\bf r}_0), ~~\mbox{and}~~~
\hat{B}= f_\sigma (\hat{\bf r}-{\bf r}_1),
\end{equation}
where $f_\sigma({\bf r})$ is a Gaussian function of width $\sigma$:
\begin{equation}
f_\sigma({\bf r}) = \frac{1}{( 2 \pi \sigma)^{d/2}}
\exp\left\{ -\frac{{\bf r}^2}{2 \sigma}  \right\}.
\end{equation}
A straightforward calculation of the Wigner representations yields 
\begin{eqnarray}
[AB]_d({\bf x'},{\bf x}) = 
f_\sigma({\bf r}-{\bf r}_0)f_\sigma({\bf r}'-{\bf r}_1), 
\end{eqnarray}
and 
\begin{eqnarray}
[AB]_{s,c}({\bf x'},{\bf x}) = h^d \delta ({\bf r}- {\bf r'})
f_\sigma ({\bf r}-{\bf r}_0)f_\sigma ({\bf r}'-{\bf r}_1) \nonumber \\ 
\times f_{\frac{\hbar^2}{2\sigma}}({\bf p}\! \pm\! {\bf p}')
\exp \left\{ \frac{({\bf r}_1 \!-\! {\bf r}_0)^2}{4 \sigma} \!+\!
 \frac{ i}{\hbar}
({\bf p \! \pm \! p'}) ({\bf r}_1 \!-\! {\bf r}_0)\right\} ,
\end{eqnarray} 
where the $+$ and $-$ signs of ${\bf p}'$ correspond to 
$[AB]_{c}$ and $[AB]_{s}$
respectively. $[AB]_d({\bf x'},{\bf x})$ is independent of the
momenta and large when ${\bf r} \sim {\bf r}_0$ and
${\bf r}' \sim {\bf r}_1$. On the other hand, 
$[AB]_{s,c}({\bf x'},{\bf x})$ are exponentially
small for all values of ${\bf x}$ and ${\bf x'}$,
when  $|{\bf r}_0$ - ${\bf r}_1| \gg \sigma $.
If ${\bf r}_0={\bf r}_1$ and $\sigma |{\bf p}|^2/\hbar^2 \gg 1$, 
then  $[AB]_{s,c}({\bf x'},{\bf x}) \approx
h^d f_\sigma^2({\bf r}-{\bf r}_0) \delta({\bf r'}-{\bf r})
\delta({\bf p}\pm{\bf p'}) $

As a second example consider the case 
\begin{eqnarray}
\hat{A}=\hat{B}=|\phi \rangle \langle \phi |.
\end{eqnarray}
Here both observables equal to the same projection operator.
Assuming the wave function $\phi({\bf r})$
to be real (as for the case of time reversal symmetry) one immediately
sees that all types of Wigner transforms are precisely the same: 
\begin{eqnarray}
[AA]_{s,c}({\bf x}',{\bf x})=[AA]_{d}({\bf x}',{\bf x})= 
 \rho_\phi({\bf x}')\rho_\phi({\bf x}), \label{AAW}
\end{eqnarray}
where 
\begin{equation}
\rho_\phi ({\bf x}) = 
\int d{\bf q} e^{\frac{i}{\hbar}{\bf p \cdot q}} 
\langle {\bf r}- \frac{\bf q}{2} |\phi \rangle \langle \phi 
|{\bf r}+ \frac{\bf q}{2}\rangle  \label{Wignerfunction}  
\end{equation}
is the Wigner function of the state $\phi({\bf r})$. $\rho_\phi ({\bf x})$
is a real function which upon small smearing 
in phase space yields a positive definite function that can be 
interpreted a ``classical'' density distribution\cite{Berry77}.

\subsection{One-point functions}

To begin with the calculation of correlation functions,
it is instructive to consider, first, the simplest case of  
one-point functions:
\begin{equation}
C_A= \left\langle \mbox{Tr}\left\{ \hat{A}~\Im G^-(\epsilon) \right\} 
\right\rangle
= \int \frac{d{\bf x}}{h^d}
  A({\bf x})\langle \Im G^-({\bf x})\rangle.  \label{1point}
\end{equation}
Here $\hat{A}$ is an observable, and $\Im G^-(\epsilon)$ is the imaginary
part of the retarded Green function. The second equality in the above
formula is obtained by substituting (\ref{Iwigner}) into the trace,
and taking into account that energy averaging acts only 
on the Green function. This averaging leaves only the Weyl 
contribution (\ref{Weyl}), therefore
\begin{equation}
C_A=  \pi \langle \langle A({\bf x}) \rangle \rangle , \label{onepoint}
\end{equation}
where double brackets denote the microcanonical average over the energy 
shell $\epsilon={\cal H} ({\bf x})$, i.e.
\begin{equation}
\langle \langle A({\bf x}) \rangle \rangle = 
\int \frac{d{\bf x}}{h^d}  A({\bf x})
\delta(\epsilon - {\cal H} ({\bf x})). \label{mca}
\end{equation}
Notice that $\hat{A}$ and $\langle \langle A({\bf x}) 
\rangle \rangle$ do not have the same dimensions, they differ by
dimensions of energy. 

\subsection{Two-point functions}
We turn, now, to the calculation of the two-point function,
\begin{equation}
C_{AB}(\omega)=  \left\langle 
\mbox{Tr}\left\{ G^+(\epsilon+\hbar \omega) \hat{A}
\right\}
 \mbox{Tr} \left\{ G^-(\epsilon) \hat{B} \right\} \right\rangle_c , 
\label{2point}
\end{equation}
where $\langle \cdots \rangle_c$ denotes the connected part of
the correlator\cite{comment1}.  Within the semiclassical
approximation it means that only the oscillatory parts
of the Green functions contribute to $C_{AB}(\omega)$. 
Inserting complete
sets of position states one can write the correlator (\ref{2point}) as
\begin{eqnarray}
C_{AB}(\omega)= \int d{\bf r}_0 d{\bf r}_1 d{\bf r}_2 d{\bf r}_3
 A({\bf r}_0,{\bf r}_1)
B({\bf r}_2,{\bf r}_3) \nonumber \\
 \times \langle G^+_{osc}({\bf r}_1,{\bf r}_0;\epsilon+\hbar\omega) 
G^-_{osc}({\bf r}_3,{\bf r}_2;\epsilon) \rangle,~~~~~~~ \label{cabo}
\end{eqnarray}
where
\begin{eqnarray}
A({\bf r}_0,{\bf r}_1)= \langle {\bf r}_0|\hat{A}|{\bf r}_1 \rangle,~~
B({\bf r}_2,{\bf r}_3)= \langle {\bf r}_2|\hat{B}|{\bf r}_3 \rangle.
\end{eqnarray}
Next, we evaluate the average of $\langle G^+_{osc}G^-_{osc}\rangle$. 
For this purpose notice that, to obtain
a nonzero contribution, one has to pair orbits of the two Green function with
similar actions. This condition imposes restrictions
as for the possible configuration of the coordinates ${\bf r}_0, {\bf r}_1,
{\bf r}_2$, and ${\bf r}_3$. 
The possibilities (in systems with time 
reversal symmetries) are: (i) 
${\bf r}_0 \sim {\bf r}_3$ and ${\bf r}_1 \sim {\bf r}_2$, as illustrated 
in Fig.~2a;
(ii) ${\bf r}_0 \sim {\bf r}_2$ and ${\bf r}_1 \sim {\bf r}_3$, see Fig 2b;
(iii) two possibilities corresponding to the case ${\bf r}_0 \sim {\bf r}_1$ 
and ${\bf r}_2 \sim {\bf r}_3$, where the orbits are deformations
of periodic orbits, see Figs. 2c and 2d.
These pairing possibilities imply that the two-point 
function is a sum of terms:
\begin{eqnarray}
C_{AB}(\omega)= C_{AB}^a(\omega)+C_{AB}^b(\omega)+C_{AB}^{c+d}(\omega),
\label{cabsum}
\end{eqnarray}
where the superscripts $a, b, c$ and $d$ refer to the contributions
associated with configurations of trajectories as shown in Fig.~2.
{\narrowtext
\begin{figure}[h]
\epsfxsize=8.5cm
%\vspace{-0.5 cm}
\epsfbox{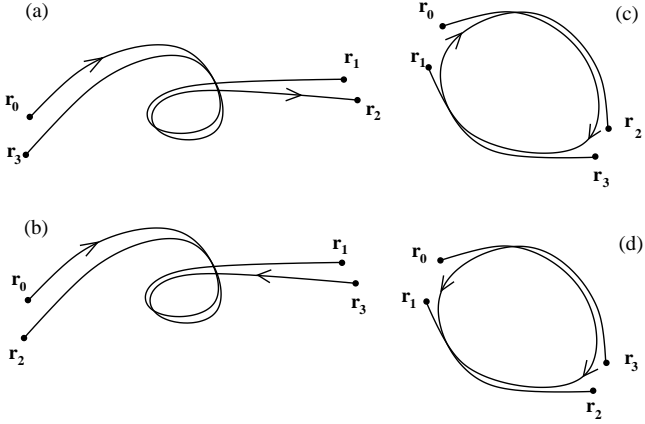}
\vspace{0.5 cm}
\caption{The orbit configurations contributing to the average of the
 Green functions product $\langle G^+_{osc}({\bf r}_1,{\bf r}_0) G^-_{osc}
({\bf r}_3,{\bf r}_2) \rangle$.} 
\label{fig:2}
\end{figure}
}

We turn to calculate the various terms of $C_{AB}(\omega)$. 
Consider, first, $C_{AB}^a(\omega)$
associated with the configuration of orbits  shown 
Fig.~2a. We start by substituting  the Wigner 
representation (\ref{Wigners}), implied by the pairing of initial and 
final points, into (\ref{cabo}). Then we change variables to 
\begin{eqnarray}
\begin{array}{ccc}
{\bf r} &=& ({\bf r}_0+{\bf r}_3)/2 , \\
{\bf q} &=& {\bf r}_3-{\bf r}_0, 
\end{array}~~~~~~
\begin{array}{ccc}
{\bf r'}&=& ({\bf r}_1+{\bf r}_2)/2, \\  
 {\bf q'} &=&{\bf r}_1-{\bf r}_2,  
\end{array} \nonumber
\end{eqnarray}
and perform the average over the Green functions. This average gives
precisely the classical propagator $\Pi({\bf x'},{\bf x};\omega)$
(see Eq.~\ref{pi1}), therefore
\begin{eqnarray}
C_{AB}^a(\omega)&=& \int d{\bf x} d{\bf x'}  [AB]_s({\bf x},{\bf x}')
\Pi({\bf x'},{\bf x};\omega) \label{caba}\\
& =&   
\frac{ 2 \pi}{\hbar}   \int_0^\infty dt ~e^{i \omega t}
\langle \langle ~[AB]_s({\bf x},{\bf x}(t)) ~\rangle \rangle, \nonumber
\end{eqnarray}
where in the second line we substitute formula (\ref{pi3}) for  
$\Pi({\bf x'},{\bf x};\omega)$, and integrate over ${\bf x}'$.

The calculation of the contribution, coming from orbits of the 
type shown in Fig.~2b, follows along the same lines. The difference is that 
now we use the second Wigner representation, Eq.~(\ref{Wignerc}). 
Changing, again, variables to means and differences of the initial 
coordinates, and using the time reversal symmetry of the system, 
$G^\pm({\bf r}_3,{\bf r}_2)=
G^\pm({\bf r}_2,{\bf r}_3)$,  yields
\begin{eqnarray}
C_{AB}^b(\omega)&=& \int d{\bf x} d{\bf x'}  [AB]_c({\bf x},{\bf x'})
\Pi({\bf x'},{\bf x};\omega) \label{cabb}\\
& =&   
\frac{ 2 \pi}{\hbar}  \int_0^\infty dt ~e^{i \omega t}
\langle \langle ~[AB]_c({\bf x},{\bf x}(t)) ~\rangle \rangle. \nonumber
\end{eqnarray}

The computation of the last two contributions, associated with the orbits
of Figs.~2c and 2d, requires an additional step. This step comes in order
to impose the condition that the two orbits are deformations of 
the same periodic orbit. It is achieved by 
requiring that the orbit from ${\bf r}_0$ to ${\bf r}_1$  passes near
$({\bf r}_2+{\bf r}_3)/2$, and similarly the orbit from 
 ${\bf r}_3$ to ${\bf r}_2$  passes in the vicinity of 
$({\bf r}_0+{\bf r}_1)/2$. To implement this condition it will be 
convenient to introduce a local coordinate system 
${\bf r}=(\tau, {\bf r}_\perp)$, 
where $\tau$ is the time along the trajectory while the coordinate
${\bf r}_\perp$ is perpendicular to it. Then the oscillatory part of 
the Green function satisfies
the semiclassical product relation:
\begin{eqnarray}
G^\pm_{osc}({\bf r}_1,{\bf r}_0;\epsilon)\simeq \hbar 
\!\!\int \!\!\dot{r} 
~d{\bf r}_\perp
G^\pm_{osc}({\bf r}_1,{\bf r};\epsilon)G^\pm_{osc}({\bf r},{\bf r}_0;\epsilon)
\label{Gproduct} \end{eqnarray}
This relation can be proved by calculating the integral in the stationary 
phase approximation, see Appendix C.

Substituting (\ref{Gproduct}) in (\ref{cabo}) and representing the
matrix elements of the operators $\hat{A}$ and $\hat{B}$ in terms of their
inverse Wigner transforms (\ref{Iwigner}) we obtain
\begin{eqnarray}
C_{AB}^{c+d}(\omega)= \int 
d {\bf x}  d {\bf x'} [AB]_d({\bf x},{\bf x'})
 K({\bf x'},{\bf x}),
 \end{eqnarray}
where 
\begin{eqnarray} 
K({\bf x'},{\bf x})=\frac{\hbar^2}{h^{2d}} \int\!\! 
d{\bf q} d {\bf q'} d {\bf r}_{\perp} d 
{\bf r}_{\perp}' \dot{r} \dot{r}'
e^{-\frac{i}{\hbar}({\bf p}\cdot {\bf q} +
{\bf p'}\cdot {\bf q'})} \times \nonumber \\
 \left\langle \!G^+_{osc}\!({\bf r}\!+\!\frac{{\bf q}}{2},
{\bf r'}\!+\! {\bf r}_{\perp}'; \epsilon \!+\! \hbar \omega)
G^+_{osc}\!({\bf r'}\!+\! {\bf r}_{\perp}', 
{\bf r}\!-\! \frac{{\bf q}}{2}; 
\epsilon \!+\! \hbar \omega) \right. \nonumber  \\
\times \left. G^-_{osc}({\bf r'}\!+\! \frac{{\bf q'}}{2},
{\bf r}\!+\! {\bf r}_{\perp}; \epsilon )G^-_{osc}
({\bf r}\!+\! {\bf r}_{\perp}, {\bf r'}\!-\!
\frac{{\bf q'}}{2}; \epsilon)
\right\rangle, 
 \nonumber
\end{eqnarray}
${\bf x}=({\bf r},{\bf p})$, and ${\bf x'}=({\bf r'},{\bf p'})$.
In obtaining the above integral we also changed variables to 
\begin{eqnarray}
\begin{array}{ccc}
{\bf r} &=& ({\bf r}_0+{\bf r}_1)/2 , \\
{\bf q} &=& {\bf r}_1-{\bf r}_0, 
\end{array}~~~~~~
\begin{array}{ccc}
{\bf r'}&=& ({\bf r}_2+{\bf r}_3)/2, \\  
 {\bf q'} &=&{\bf r}_3-{\bf r}_2.  
\end{array} \nonumber
\end{eqnarray}
Now, we approximate the average of the four Green functions
by the diagonal approximation. Diagonal sums in this case can be
formed in two ways coming from the two possibilities of pairing the
retarded and the advanced Green functions. These two possibilities
correspond to the orbits configurations illustrated in
Figs.~2c and 2d. The two contributions are identical
due to the time reversal symmetry of
our system. As in the calculation of the classical propagator, we
expand the actions to linear order around 
${\bf r},  {\bf r'}$, and $\epsilon$, and integrate over
${\bf q}, {\bf q'}, {\bf r}_\perp$, and 
${\bf r'}_\perp$. The result is 
\begin{eqnarray}
K({\bf x'},{\bf x}) &=& 8 \pi^2 h^{2d} \dot{r}\dot{r}'
\sum_{\mu \nu} |A_\mu|^2 |A_{\nu}|^2 e^{i \omega (t_\mu + t_{\nu})} 
\nonumber \\ &\times &
 \delta ({\bf p} - \frac{{\bf p}_\mu + {\bf p}_\nu}{2})
\delta ({\bf p}_{\mu \perp}-{\bf p}_{\nu \perp}) \nonumber \\
 & \times &
 \delta ({\bf p'} - \frac{{\bf p}_{\mu}' + {\bf p}_{\nu}'}{2})  
\delta ({\bf p}_{\mu \perp}'-{\bf p}_{\nu \perp}'), \nonumber
\end{eqnarray}
 where $t_\mu$ and $t_{\nu}$ are the periods of the orbits
going from ${\bf r}$ to ${\bf r'}$ and from ${\bf r'}$ 
to ${\bf r}$ respectively. ${\bf p}_{\mu}$ and ${\bf p}_{\nu}$ denote 
the momenta of the orbits at ${\bf r}$, and similarly 
${\bf p}_{\mu}'$ and ${\bf p}_{\nu}'$ are the momenta at
${\bf r'}$. We denote by subscript $\perp$ the perpendicular 
components of momenta that are conjugate to ${\bf r}_\perp$.   
Finally, we replace the sum over orbits by an 
integral using the sum rule (\ref{sumrule}), and obtain:
\begin{eqnarray}
C_{AB}^{c+d}(\omega)= 2 \int \frac{d t d t'}{\hbar^2} 
e^{i \omega (t+t')}
\langle \langle [AB]_d({\bf x},{\bf x}') \rangle \rangle_{t,t'}. \label{c+d}
\end{eqnarray}
The average $\langle \langle \cdots \rangle \rangle_{t,t'}$ of a general
function, $g({\bf x},{\bf x'})$, is defined as
\begin{eqnarray}
\langle \langle g({\bf x},{\bf x'}) \rangle \rangle_{t,t'}& =&
\int d{\bf x} d {\bf x'} g({\bf x},{\bf x'})
\delta (\epsilon \!-\!{\cal H}({\bf x})) 
\delta (\epsilon \!- \!{\cal H}({\bf x}')) \nonumber \\
 &\times &
\delta ({\bf x}_{ \parallel}' \!- \!{\bf x}_{ \parallel}(t))
\delta ({\bf x}_{ \parallel} \! -\! {\bf x}_{ \parallel}'(t')), 
\end{eqnarray}
where  ${\bf x}_\parallel$ denote a coordinate on the energy 
shell $\epsilon ={\cal H}({\bf x})$. The two $\delta$-functions
in the above integral
imply that the average is over periodic orbits on the energy shell
with period $t+t'$. The factor of two in (\ref{c+d}) is
due to the time reversal symmetry. 

The final result for the two-point function, $C_{AB}(\omega)$,
is the sum  (\ref{cabsum}) of 
(\ref{caba}), (\ref{cabb}), and (\ref{c+d}).
It expresses a quantum mechanical quantity, in terms 
of classical correlation functions associated with the Wigner representations
of the operators $\hat{A}$ and $\hat{B}$. There are two sources of
contributions to $C_{AB}(\omega)$: open trajectories, and periodic orbits.
The magnitude of the corresponding terms depends on properties of the
various Wigner
representations $[AB]_d({\bf x},{\bf x}')$, $[AB]_s({\bf x},{\bf x}')$,
and $[AB]_c({\bf x},{\bf x}')$. When $A({\bf x})$ and $B({\bf x}')$ are very
smooth functions, the main contribution comes from periodic orbits. If, on the
other hand, $\hat{A}$ and $\hat{B}$ equal to the same projector, the main 
contribution comes from open orbits.

\subsection{Diagrams rules for $n$-point functions in systems with 
time reversal symmetry}

The results for the two-point function can be generalized 
to $n$-point functions in systems with time reversal symmetry. 
It is instructive
to formulate this generalization in terms of a set of
rules for a diagrammatic calculation.
Only the connected part of the correlation functions
will be considered here. 
It implies that the relevant part of the Green function
is the oscillatory contribution (\ref{osc}). 

We define the dimensionless propagator
on the energy surface as
\begin{eqnarray}
\Pi_\epsilon ({\bf x}_{\parallel}',{\bf x}_{\parallel};\omega) =
\frac{2 \pi}{\hbar}
\int_0^\infty \!\!\! dt~ e^{i \omega t} }{ 
h^d \delta ({\bf x}_{\parallel}'\!-\!{\bf x}_{\parallel}(t)), 
\label{energyshell}
\end{eqnarray} 
where both ${\bf x}_{\parallel}$ and ${\bf x}_{\parallel}'$ lie on
the energy shell $\epsilon={\cal H}({\bf x})$. In order to shorten our 
notations, from now on, we omit the subscript $\parallel$. 

The rules for calculating
the average of an $n$-point functions are the following:
\begin{itemize}
\item 
Write the correlation function as a space integral involving the
matrix elements of the operators and the Green functions.
\item
Find all possible pairs of initial and final points connected
by Green functions, and construct the orbits configurations as in
Fig.~2. To each coordinate pair assign a 
phase space point.
\item 
Express the Matrix elements of the operators 
as an inverse transform of the Wigner representation
implied by the pairing configuration of the initial and final points. 
These  Wigner representation are functions of  the phases space points 
assigned in the previous step.
\item
Phase space points connected by retarded and advanced 
Green functions are
associated with the classical propagator  (\ref{energyshell})
connecting the two points.
\item
In case where two classical propagators join at the same 
phase space point, then such a point is accompanied by 
factor of  $1/2\pi$ (this factor emerges from expansions like 
(\ref{Gproduct})). 
\item
Integrate over all phase space coordinates with the weight
$\int d{\bf x} \delta(\epsilon -{\cal H}({\bf x})) (\cdots )/ h^d$.
This integration is the same as microcanonical averaging (\ref{mca}).
\end{itemize} 

The diagrams of the two-point correlation function
$C_{AB}(\omega)$ are shown in Fig.~3. The contributions $C^a_{AB}(\omega)$
and $C^b_{AB}(\omega)$ are represented by Fig.~3a and Fig.~3b, respectively.
These terms came from orbits configurations shown in Figs.~2a and 2b.
The two other contributions (Figs.~2c and 2d), 
$C^{c+d}_{AB}(\omega)$, are represented by the diagram of Fig.~3c.     

In Fig.~4 we show examples of diagrams contributing to the three-point 
correlation function,
\begin{eqnarray}
C_{ABC}(\omega_1,\omega_2)= \left\langle \mbox{Tr}
\left\{ G^+(\epsilon\!+\!\hbar \omega_1) 
\hat{A}\right\} \right.~~~~~~~~~~~
\label{3point} \\
\times \left. \mbox{Tr} \left\{ G^-(\epsilon\!+\!\hbar \omega_2) 
\hat{B}\right\} 
\mbox{Tr}\left\{ G^-(\epsilon) \hat{C} \right\}\right\rangle_c. \nonumber
\end{eqnarray}
For instance, the top diagram of this figure equals to $\langle \langle
\Pi_\epsilon({\bf x}'',{\bf x}', \omega_1-\omega_2) 
\Pi_\epsilon({\bf x}',{\bf x}, \omega_1) [ABC]_\alpha({\bf x}, 
{\bf x}',{\bf x}'')
\rangle \rangle/2\pi$,
where the microcanonical average is with respect to all the 
phase space coordinates.

{\narrowtext
\begin{figure}[h]
\epsfxsize=8.5cm
%\vspace{-0.5 cm}
\begin{center}
\epsfbox{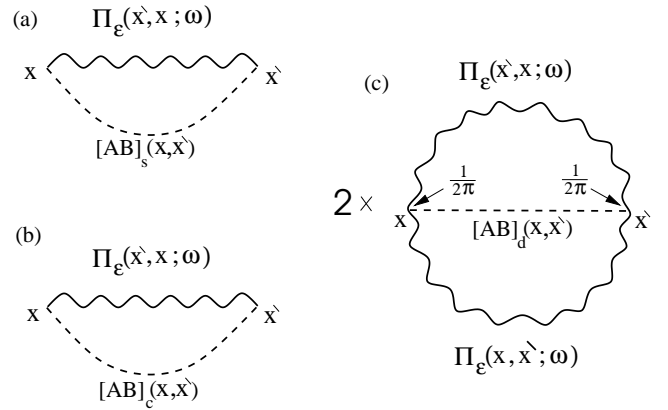}
\end{center}
\vspace{0.5 cm}
\caption{The diagrams of  
the two-point function $C_{AB}(\omega)$. Wiggly lines represent
the classical propagator on the energy shell (\ref{energyshell}), and
dashed line correspond to the vertecies given by the appropriate
Wigner representation of the observables $\hat{A}$ and $\hat{B}$.}
\label{fig:6}
\end{figure}
}

{\narrowtext
\begin{figure}[h]
\epsfxsize=8.5cm
%\vspace{-0.5 cm}
\begin{center}
\epsfbox{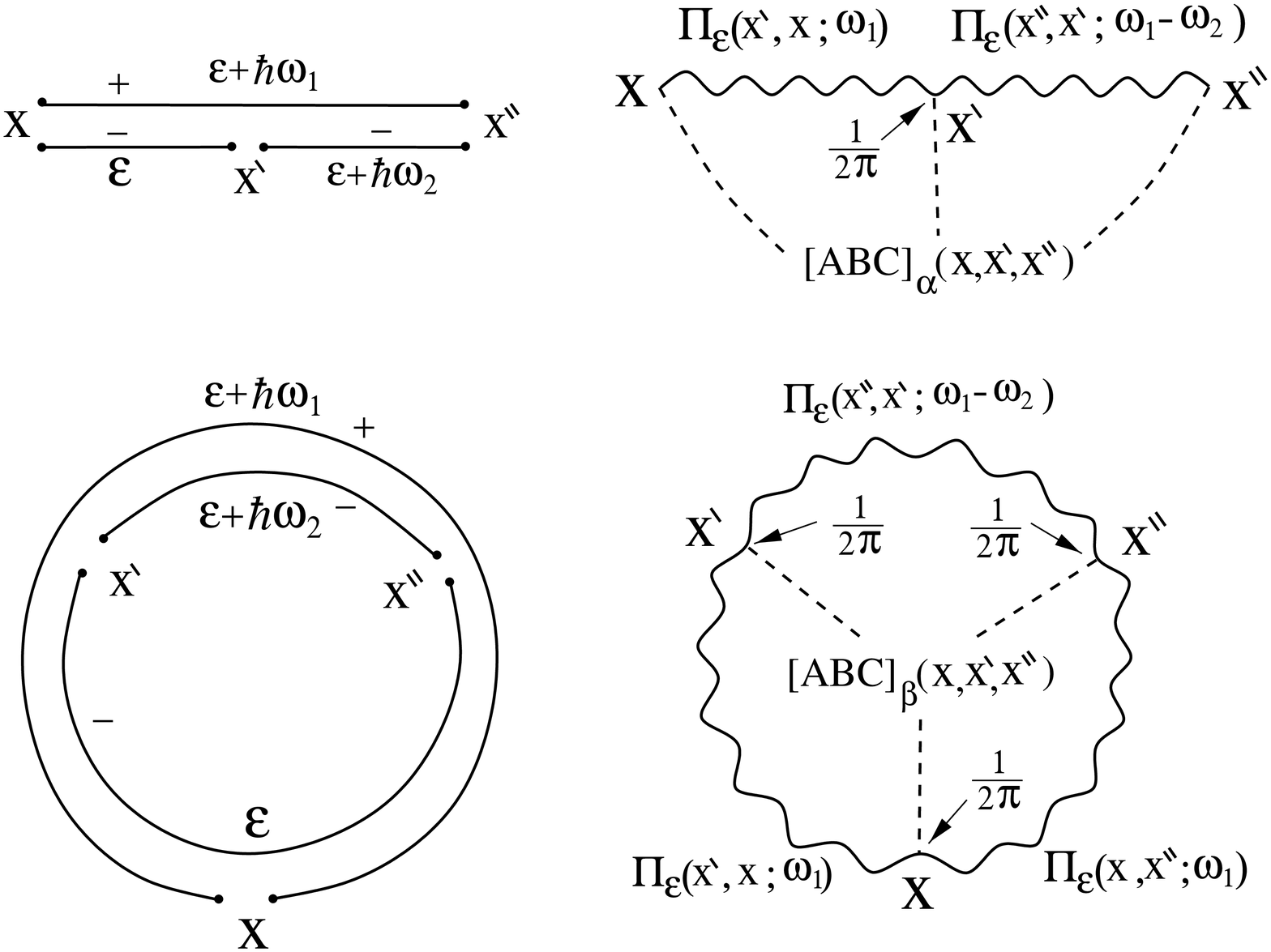}
\end{center}
\vspace{0.5 cm}
\caption{Examples of diagrams contributing to the three-point 
function (\ref{3point}). The orbits configuration corresponding
to a diagram is drawn on its left side. $[ABC]_{\alpha,\beta}({\bf x},
{\bf x'},{\bf x''})$ denote various Wigner representations of the
external product of observables, $\hat{A}\otimes \hat{B} \otimes \hat{C}$.
}
\end{figure}
}  

Finally we remark that, the procedure outlined above generates a large number
of diagrams. However, knowing some general
properties of the observables, 
such as symmetries and smoothness of their Wigner representations, can help in 
reducing the number of diagrams considerably. 

\section{Applications}
The purpose of this section is to present applications of the formalism
developed above.
In the first example we consider the $N$-disk scattering 
system\cite{Cvitanovic88,Gaspard89}.
In the second application we study the indirect photodissociation 
process of complex molecules\cite{book}. Special attention will be 
given to understand the manifestations of the individual 
imprints of the systems.
  
\subsection{The $N$-disk scattering system}
Consider a quantum particle moving in a two dimensional plain,
and scattered from $N$ disks located, say randomly near the origin, 
see Fig.~5. Resonances in this system can be pictured as situations where 
the particle is trapped for a long time in the vicinity of the disks.
A natural quantity characterizing the system is the
probability of finding it in a final state $|\psi_f \rangle$ 
when prepared in an initial state $|\psi_i \rangle$.
{\narrowtext
\begin{figure}[h]
\epsfxsize=8.2cm
%\vspace{-0.5 cm}
\epsfbox{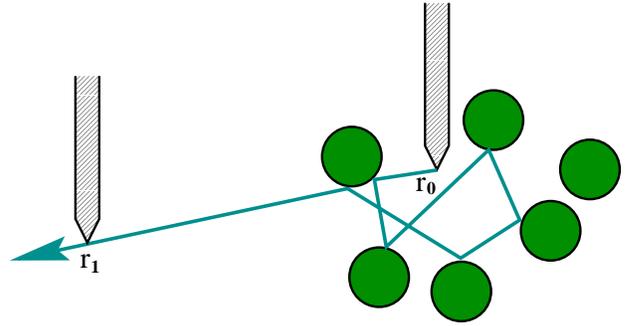}
\vspace{0.5 cm}
\caption{An illustration of the classical $N$-disk scattering system. 
In the experimental realization of the quantum analogue of this system,
using microwave cavities, a transmitor located 
at ${\bf r}_0$ excites a microwave field which is probed at
${\bf r}_1$.}  
\end{figure}
}
\noindent
In the experimental realization of this system\cite{Lu99}, using microwaves 
cavities, a transmitor prepares the particle in a localized wave packet
at ${\bf r}_0$, and a detector collects the particle 
 at  ${\bf r}_1$. The signal is proportional to
$|G^+({\bf r}_1,{\bf r}_0;\epsilon)|^2$, and its
simplest statistical characteristics
is the correlation function:
\begin{eqnarray}
K(\omega)=\frac{ \langle |G^+({\bf r}_1,{\bf r}_0;\epsilon)|^2
|G^+({\bf r}_1,{\bf r}_0;\epsilon+ \hbar\omega)|^2 \rangle}{
\langle |G^+({\bf r}_1,{\bf r}_0;\epsilon)|^2 \rangle^2}-1.
\end{eqnarray}

To implement the diagrammatic approach,
we start by writing  $|G^+({\bf r}_1,{\bf r}_0;\epsilon)|^2$ as a trace:
\begin{eqnarray} 
|G^+({\bf r}_1,{\bf r}_0;\epsilon)|^2 \!=\! \mbox{Tr}
\left\{ G^+(\epsilon) \delta (\hat{\bf r}\!-\!{\bf r}_0)
 G^-(\epsilon)  \delta (\hat{\bf r}\!-\!{\bf r}_1) \right\}. \label{disco}
\end{eqnarray}
The orbit configuration representing the above formula
is shown in Fig.~6a. Assuming the distance
$|{\bf r}_1- {\bf r}_0|$, to be larger than the particle wavelength
there will be only one contribution associated with the oscillatory
parts of the Green functions. The diagram representing this contribution
is shown in Fig.~6b. A straightforward calculation of this diagram
yields
\begin{eqnarray}
\langle |G^+({\bf r}_1,{\bf r}_0;\epsilon)|^2 \rangle =
\int_0^{\infty} \! dt ~f({\bf r}_1,{\bf r}_0;t),
\end{eqnarray}
where 
\begin{eqnarray}
f({\bf r}_1,{\bf r}_0;t) =\int \frac{d{\bf p}_{0}}{h^{d-1}\hbar^2}
\delta [\epsilon - {\cal H}({\bf x}_0)]
\delta[{\bf r}_1 - {\bf r}({\bf x}_0,t)]
\end{eqnarray} 
is proportional to
the classical probability of finding particle at ${\bf r}_1$,
starting from ${\bf r}_0$ in all possible directions, and evolving according 
to the classical equations of motion. Here ${\bf r}({\bf x}_0,t)$ denotes the
trajectory of the particle, as function of time $t$, starting from the  
initial phase space  point ${\bf x}_{0}= ({\bf r}_0,{\bf p}_0)$.

The correlator $K(\omega)$ is
proportional to the diagram shown in Fig.~6d, where again we assume
the distance between ${\bf r}_1$ and ${\bf r}_0$ to be much longer
than the particle wavelength. A straightforward calculation of this diagram
leads to
\begin{eqnarray} 
K(\omega) = \left|\frac{F({\bf r}_1,{\bf r}_0;\omega)}{
F({\bf r}_1,{\bf r}_0;0)} \right|^2, \label{komega}
\end{eqnarray}
where 
\begin{eqnarray} 
F({\bf r}_1,{\bf r}_0;\omega)=
\int_0^\infty \!\! dt~ e^{i \omega t} 
f({\bf r}_1,{\bf r}_0;t) 
\end{eqnarray}
is the Fourier transform of $f({\bf r}_1,{\bf r}_0;t)$ for time $t>0$.
{\narrowtext
\begin{figure}[h]
\epsfxsize=8.7cm
%\vspace{-0.5 cm}
\epsfbox{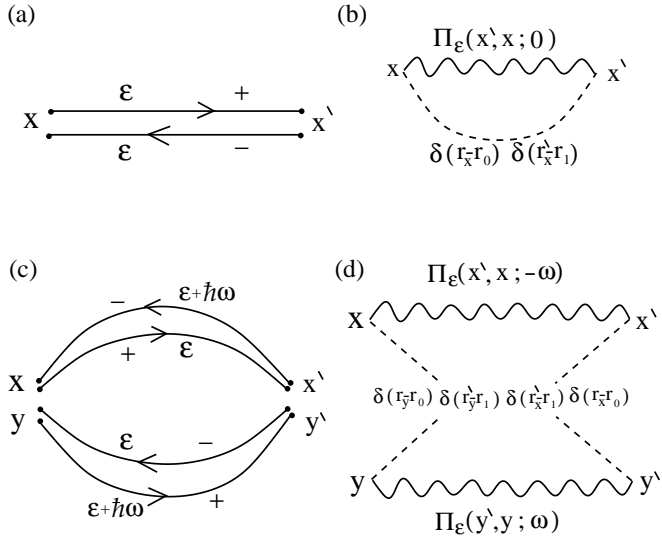}
\vspace{0.5 cm}
\caption{The diagrams of the correlation function $K(\omega)$.
(a) The orbit configuration, and (b) the corresponding diagram  
for $\langle |G^+({\bf r}_1,{\bf r}_0;\epsilon)|^2\rangle$ .
(c) The orbit configuration, and (d) diagram for the connected part of 
$\langle |G^+({\bf r}_1,{\bf r}_0;\epsilon)|^2
|G^+({\bf r}_1,{\bf r}_0;\epsilon+ \hbar\omega)|^2 \rangle$.} 
\end{figure}
}
In understanding the typical behavior of $K(\omega)$, it is necessary to
characterize the function $f({\bf r}_1,{\bf r}_0;t)$.
The spectral decomposition of classical propagator (\ref{decomposition}) 
implies that\cite{Gaspard92}  
\begin{eqnarray} 
f({\bf r}_1,{\bf r}_0;t) =\sum_\alpha b_\alpha e^{- \gamma_\alpha t}, \label{f}
\end{eqnarray}
where $\gamma_\alpha$ are the Ruelle resonances of the system, 
and $b_\alpha$ are coefficients associated with the eigenfunctions of the 
classical propagator, $\chi_\alpha^l({\bf x})$ and $\chi_\alpha^r({\bf x})$, 
as well as the precise positions of the transmitor and the detector.
The first eigenvalue $\gamma_0$ 
is the escape rate of the system. This eigenvalue dominates the behavior 
of the system at long times, i.e. small $\omega$. Taking only its contribution,
the correlation function reduces to a Lorentzian, 
$K(\omega) \approx \gamma_0^2/(\gamma_0^2+ \omega^2)$. 
This result has been known long ago in nuclear 
physics\cite{Ericsson}. It was reproduced in the context
of chaotic scattering\cite{Blumel88,Jalabert90,Lai92}, and also measured
experimentally in a microwave cavities\cite{Doron90}. 
  
We turn to consider the  nonuniversal imprints of the
system on $K(\omega)$. For this purpose we shall take into 
account contributions 
of higher Ruelle resonances in (\ref{f}).
It will be  convenient to  rescale all frequencies with $\gamma_0$,
and set $b_0=1$ (without loss of generality). 
Then, adding the next term of (\ref{f}), $F(\omega)$ takes the form
\begin{equation}
F(\omega)\simeq \frac{1}{1 + i\omega}+\sum_\pm \frac{ b_1}{\gamma'_1 + i
 (\omega \pm \gamma''_1)}, \label{nextapprox}
\end{equation}
where $\gamma'_1$ and $\gamma''_1$ denote the real and the imaginary parts 
of $\gamma_1$, respectively, and we assume $b_1$ to be real. 
In Fig.~7 we plot the correlation function $K(\omega)$ calculated 
from (\ref{komega}) using (\ref{nextapprox}). The solid line is
the Lorentzian obtained when $b_\alpha=0$ for $\alpha>0$. 
The dashed line is a representative example of the nonuniversal
behavior of the system (with $b_1=1.5,~ \gamma_1'=2, ~\gamma_1''=4$, and 
$b_\alpha=0$ for $\alpha>1$ ).

{\narrowtext
\begin{figure}[h]
\epsfxsize=8.5cm
%\vspace{-0.5 cm}
\epsfbox{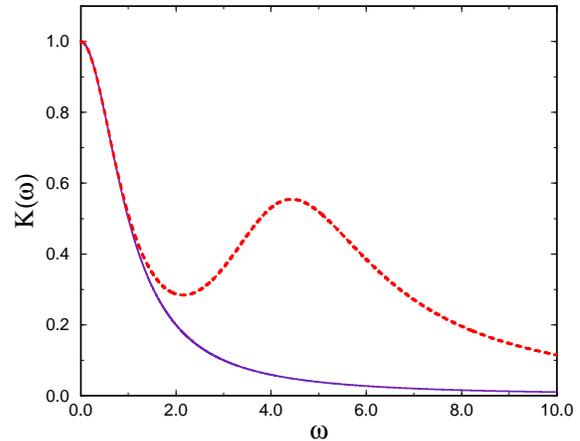}
\vspace{0.5 cm}
\caption{The correlation function $K(\omega)$. The solid line correspond
to the universal limit where $b_\alpha=0$  for all $\alpha >0$.
The dotted line includes nonuniversal imprints of the system, with
 $b_1=1.5$,  $\gamma'_1=2 $, and $\gamma''_1=4$.}
\end{figure}
}

A behavior very similar to that shown in Fig.~7 has been recently 
observed in a microwave experiment on the 4-disk system\cite{Lu99}.
Our results explain the additional peak in the measured correlation 
function, $K(\omega)$, as the imprints of the complex
Ruelle resonances of the system\cite{comment2}.

\subsection{Statistics of photodissociation spectra}

As a second application of the semiclassical diagrammatic approach,
we study the statistics of photodissociation spectra
of complex molecules, such as the radicals $HO_2$ and $NO_2$. 
Disintegration of such molecules is a two step process. In the first step, 
a photon excites the molecule to an energy above the 
dissociation threshold. Then fragmentation proceeds by
redistribution of energy in the vibrational degrees of freedom, or
tunneling from binding to unbinding energy surfaces of the adiabatic 
electronic potential\cite{book}. A barrier, which separates
quasi-stable states from continuum modes, hinders the immediate 
dissociation of the excited molecule. The large number of degrees of freedom
and the complexity of the system imply that, on these long lived resonances, 
the dynamics of the system is chaotic. An illustration of such a system is 
depicted in Fig.~8.
{\narrowtext
\begin{figure}[h]
\epsfxsize=8.2cm
%\vspace{-0.5 cm}
\epsfbox{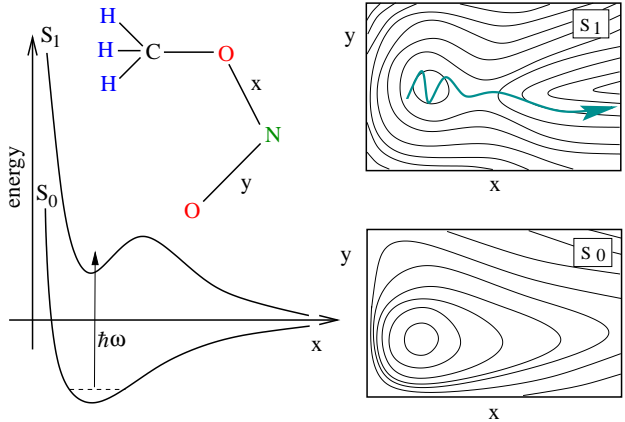}
\vspace{0.5 cm}
\caption{An illustration for the electronic energy surfaces of methyl nitrite,
CH$_3$ONO, as an example for a complex molecule which undergoes an indirect
photodissociation process. $S_0$ and $S_1$ are contour plots of the 
electronic surface potential associated with the ground and the excited 
electronic states. They are plotted as functions of the distances between 
the Nitrogen and the Oxygen atoms, 
keeping all other degrees of freedom fixed at their equilibrium values.
The irregular shape of $S_1$ leads to chaotic dynamics of the excited molecule.
} 
\end{figure}
}
Consider a molecule, in the ground state $|g\rangle$, excited
by a light pulse to an energy above  the dissociation threshold, 
and let $\hat{{\cal H}}$ denote the Hamiltonian of the system on the
excited electronic surface ($S_1$ in the illustration of Fig.~8).
It will be assumed that  $\hat{{\cal H}}$ represents an open system with
several open channels. The photodissociation cross section of the molecule, 
in the dipole approximation, is given by
\begin{equation} 
\sigma(\epsilon)= \eta~ 
\mbox{Im}~\mbox{Tr} \left\{ \hat{A} G^-(\epsilon) \right\} \label{sigma} 
\end{equation} 
where $G^-(\epsilon)$ is the retarded Green function of the molecule, 
and $\hat{A}$ is a projection operator given by 
\begin{equation}
\hat{A}= |\phi\rangle \langle \phi |,~~~\mbox{where}~~~ |\phi \rangle=  
{\cal D} ~| g \rangle . \label{hata}
\end{equation} 
Here ${\cal D} = {\bf d}\cdot \hat{\bf e}$ is the projection of the 
electronic dipole moment operator of the molecule, ${\bf d}$, on the 
polarization, $\hat{\bf e}$, of the absorbed light, and  $\eta=\epsilon/c 
\hbar \varepsilon_0$,  $c$ being the speed of light, and $\varepsilon_0$ the 
electric permitivity. The energy $\epsilon$ is measured
from the ground state of the molecule.   

A natural statistical characteristic of the photodissociation
process is the dimensionless two-point correlation function: 
\begin{equation} 
Z(\omega) =\frac{\left\langle  \sigma(\epsilon ) 
\sigma(\epsilon + \hbar \omega ) \right\rangle_c}{\langle   
\sigma(\epsilon)\rangle^2}. \label{def} 
\end{equation} 
It will be assumed that 
the excitation energy,  $\epsilon$, is  sufficiently high such  
that the mean spacing between the vibrational modes of the molecule 
is smaller than the energy broadening due to the finite life time 
of the system in the excited states. This is the regime of 
overlapping resonances.  

From (\ref{sigma}), (\ref{1point}), and (\ref{2point}) it follows that
the correlation function (\ref{def}) equals to 
\begin{eqnarray}
Z(\omega)= \frac{1}{2}\Re \frac{C_{AA}(\omega)}{C_A^2},  \label{Z}
\end{eqnarray}
where $\hat{A}$ is the projection operator (\ref{hata}), while
$C_{A}$ and $C_{AA}(\omega)$ are the corresponding one-point and
two-point functions. The special feature of the operator 
$\hat{A} \otimes \hat{A}$ is that all its
Wigner representations are identical, see Eq.~(\ref{AAW}). 
It implies that (\ref{Z})  has the form
\begin{eqnarray}
Z(\omega) \simeq Z_1(\omega)+ Z_2(\omega), \label{Zsum}
\end{eqnarray}
where $Z_1(\omega)$ is the contribution of open orbits (Figs.~2a and 2b),
\begin{eqnarray}
Z_1(\omega) = \frac{2}{\pi \hbar} \Re\int_0^\infty \!\!\! d t ~ e^{i \omega t}~
\frac{\langle \langle  \rho_\phi({\bf x}(t))\rho_\phi({\bf x}) \rangle 
\rangle}{\langle \langle  \rho_\phi({\bf x})\rangle \rangle^2}, \label{Z1}
\end{eqnarray}
while $Z_2(\omega)$ is the contribution of periodic orbits 
(Figs.~2c and 2d),
\begin{eqnarray}
Z_2(\omega) = \frac{1}{\pi^2 \hbar^2} \Re \int_0^\infty \!\!\! d t dt'~ 
e^{i \omega (t+t')} 
\frac{\langle \langle \rho_\phi({\bf x})\rho_\phi({\bf x}') 
\rangle \rangle_{t,t'}}{\langle \langle  
\rho_\phi({\bf x})\rangle \rangle^2}. \label{Z2}
\end{eqnarray}
$\rho_\phi({\bf x})$ in the above formulae is the Wigner function 
of the initial state ${\cal D}|g \rangle$.

Consider the limit of small $\omega$. This limit reflects the
long time behavior of the correlation functions
$\langle \langle\rho_\phi({\bf x}(t))\rho_\phi({\bf x}) \rangle \rangle$
and $\langle \langle\rho_\phi({\bf x})\rho_\phi({\bf x'}) \rangle 
\rangle_{t,t'}$. The leading contribution, in this case, comes from
the smallest eigenvalue in the spectral decomposition (\ref{decomposition}), 
i.e. $\gamma_0$. Taking into account only this eigenvalue we obtain
\begin{eqnarray}
\langle \langle\rho_\phi({\bf x}(t))\rho_\phi({\bf x}) \rangle \rangle
& \simeq & a e^{-\gamma_0 t}, \nonumber \\
\langle \langle\rho_\phi({\bf x})\rho_\phi({\bf x'}) \rangle 
\rangle_{t,t'} & \simeq & b e^{-\gamma_0 (t+t')}, \nonumber
\end{eqnarray}
where 
\begin{eqnarray}
a&=& h^d \langle \langle \rho_\phi({\bf x})\chi^l_0({\bf x}) \rangle \rangle
\langle \langle \rho_\phi({\bf x})\chi^r_0({\bf x}) \rangle \rangle, \label{a}
\\ b &=& h^{2d}  \langle \langle \rho_\phi({\bf x})\chi^l_0({\bf x})
\chi^r_0({\bf x}) \rangle \rangle^2. \label{b}
\end{eqnarray}
The ratio $b/a=\Delta$ has an energy dimension. In the semiclassical 
limit, $\hbar \to 0$, it approaches zero as $\hbar^d$, thus,
it corresponds  to the mean spacing between resonances.
Substituting the above approximations of the classical correlation 
functions into (\ref{Z1}) and (\ref{Z2}), and rescaling the energies 
as  $\hbar \omega = \Omega \Delta$ and  $\hbar \gamma_0 =  \Gamma \Delta$,
one obtains
\begin{eqnarray}
Z(\Omega) \approx Z^{(0)}(\Omega)= 
\frac{2 \xi}{\pi} \left(\frac{\Gamma}{\Gamma^2+\Omega^2}+ 
\frac{1}{2\pi}
\frac{\Gamma^2-\Omega^2}{(\Gamma^2+\Omega^2)^2} \right),  \label{Zuniversal}
\end{eqnarray}
where $\xi= (a/\langle \langle \rho_\phi \rangle \rangle)^2/b$ is a 
dimensionless constant of order unity. The first term in the brackets comes
from $Z_1(\omega)$, i.e. from open trajectories of the type
shown in Figs.~2a and 2b. This term constitutes the leading contribution
to $Z(\omega)$ when  $\Gamma > 1$.  In the limit  $\Gamma \ll 1$, the 
second term in the brackets becomes dominant. This term, associated with 
$Z_2(\omega)$, comes from the periodic orbits of the system.
Notice, however, that  the regime $\Gamma \ll 1$, corresponding to less than 
one open channel, is beyond our semiclassical approximation.

Equation (\ref{Zuniversal}) represents the universal limit of the correlator
$Z(\Omega)$. It was first derived by Fyodorov and 
Alhassid\cite{Fyodorov98} using the nonlinear $\sigma$-model\cite{Efetov83}.
Our derivation confirms their conjecture that, in the limit of overlapping 
resonances ($\Gamma >1$), $Z(\omega)$ can be derived by semiclassical methods. 
An alternative derivation of (\ref{Zuniversal}), using random matrix theory,
is presented in Appendix D.

Yet, the range of validity of formulae (\ref{Zsum}-\ref{Z2}) goes far beyond 
the universal regime. They account also for system 
specific contributions which contain a valuable information
about the nature of the system. This individual imprints 
come from the higher eigenvalues 
and eigenvectors of the classical propagator (\ref{decomposition}).

For example, the leading nonuniversal contribution coming from 
$Z_1(\omega)$ is given by
\begin{eqnarray}
 Z_1^{(1)}(\Omega) \simeq \frac{2 \xi a_1}{\pi} \sum_\pm 
\frac{\Gamma_1}{\Gamma_1^2+(\Omega\pm\Omega_1)^2} . \nonumber
\end{eqnarray}
Here $\Gamma_1= \hbar \gamma_1' /\Delta$ and 
$\Omega_1=  \hbar \gamma_1''/\Delta$, where $\gamma_1'$ and  $\gamma_1''$
are the real and the imaginary parts of the second Ruelle resonance. 
The constant
$a_1$ comes from an integral similar to (\ref{a}) but with the eigenfunctions 
$\chi_1^{l,r}({\bf x})$. 

Similarly, the leading nonuniversal contribution of $Z_2(\Omega)$ is
\begin{eqnarray}
Z_2^{(1)}(\Omega) \simeq \frac{2 \xi b_{01}}{\pi^2} \sum_\pm 
\frac{\Gamma \Gamma_1-\Omega (\Omega\pm\Omega_1)}
{(\Gamma^2+ \Omega^2)(\Gamma_1^2+(\Omega\pm\Omega_1)^2)}, \nonumber
\end{eqnarray}
where $b_{01}$ is a constant depending on $\rho_\phi({\bf x})$ and 
the eigenfunctions $\chi_0^{l,r}({\bf x})$ and $\chi_1^{l,r}({\bf x})$.

The two-point function $Z(\Omega)$, in the approximation
\begin{eqnarray}
Z(\Omega)\simeq Z^{(0)}(\Omega)+ Z_1^{(1)}(\Omega)+ Z_2^{(1)}(\Omega),
\label{Znon}
\end{eqnarray}
is plotted in Fig.~9. The solid line 
represents the universal form $Z^{(0)}(\Omega)$ given 
by Eq.~(\ref{Zuniversal}). 
The dotted line
shows the typical behavior of systems where decay of correlations 
is of diffusive nature. It appears when the subleading Ruelle resonances
are purely real ($\Omega_1= 0$). 
In this case the deviation from $Z^{(0)}(\Omega)$ 
appears mainly as an increase of correlations near $\Omega =0$. 
The dashed line corresponds to the case of a complex Ruelle resonance 
($\Omega_1 \neq 0$). It characterizes the behavior of ballistic systems. 
Here the nonuniversal contribution is located in the tail of $Z(\Omega)$,
near $\Omega=\Omega_1$, where the universal term (\ref{Zuniversal}) is 
already negligible. These plots demonstrate the significance
of the individual imprints of the system on $Z(\Omega)$. 

{\narrowtext
\begin{figure}[h]
\epsfxsize=8.5cm
%\vspace{-0.5 cm}
\epsfbox{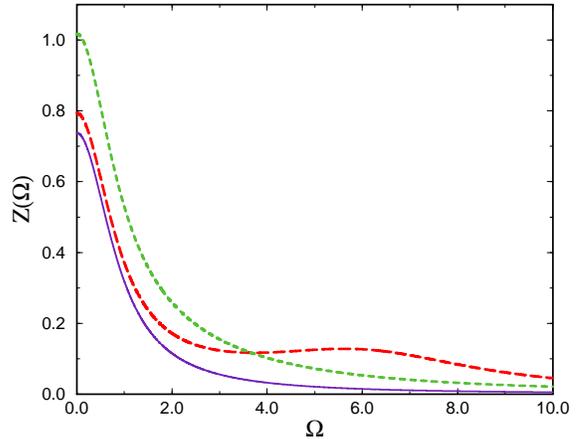}
\vspace{0.5 cm}
\caption{ The two-point function, $Z(\Omega)$, 
of the photodissociation cross section for systems with time reversal symmetry.
The solid line is the universal result (Eq.~\ref{Zuniversal}) 
for the case $\xi=\Gamma = 1$. The dashed and the dotted lines show
the two-point function which includes system specific 
contributions in the approximation of Eq.~(\ref{Znon}). In both cases 
$\Gamma=1$, $\Gamma_1=3$ and $a_1=b_{01}=1$, but, the dashed line 
corresponds to $\Omega_1=6$, while the dotted line to $\Omega_1=0$.}
\end{figure}
}
%%%%%%%%%%%%%%%%%%%%%%%%%%%%%%%%%%%%%%%%%%%%%%%%%%%%%%%%%%%%%%%%%%%%%%%%%%
\section{Summary}

The main result of this paper is the construction of diagrammatic approach for 
calculating the $n$-point functions of open chaotic systems with
time reversal symmetry. In essence, this approach is simply an economic way for
evaluating the ``diagonal approximation'' for the $n$-point functions. 
The result is expressed in terms of classical correlation functions associated
with the observables. These correlation functions reflect properties of the 
Perron-Frobenius spectral decomposition of the system, i.e. features of the
irreversible dynamics of probability densities on the energy shell.
An important ingredient of this formalism is the generalization of
Wigner transforms for external products of operators. 

The strength of this formalism is in its capability of treating particular 
systems rather than ensemble of them. Nevertheless, in the appropriate limit,
our results reduce to those of random matrix theory. This was demonstrated by
considering two examples: The $N$-disk scattering system, and the process of
indirect photodissociation in complex molecules. The weakness of this 
diagrammatic approach is the lack of a clear systematic procedure for 
calculating quantum corrections 
such as weak localization correction. This limits our treatment to systems 
with escape rate sufficiently large so that the system disintegrate
before weak localization effects set in.  

The focus of this paper was on systems with time reversal 
symmetry, and no other discrete symmetries. Within the universal limit,
a generalization of our results to system without time reversal symmetry is
straightforward. However, when considering nonuniversal manifestations, 
the situation is more complicated. The reason is that, usually in ballistic 
systems, there is no separation between the time scales on which
nonuniversal features appear, and that where time reversal symmetry is broken. 

Further development of this diagrammatic approach
should include a prescription for calculating
weak localization corrections, generalization to systems without time 
reversal symmetry, and to systems with other  discrete symmetries, such as 
reflection or inversion. 
\newline

\noindent
{\bf Acknowledgments}
\newline

\noindent
This work was initiated during the workshop on {\em Dynamics of Complex
 Systems}
which took place at  the Max Plank Institute in Dresden, 1999.
I thank the organizers and the institute for the generous hospitality.
\newline
%%%%%%%%%%%%%%%%%%%%%%%%%%%%%%%%%%%%%%%%%%%%%%%%%%%%%%%%%%%%%%%%%%%%%
\appendix{\bf Appendix A: The local  term, 
$\Pi_{loc}({\bf x}',{\bf x};\omega))$.}
\newline

In this appendix we calculate the Weyl contribution to the classical 
propagator, 
$\Pi({\bf x'},{\bf x};\omega)$.  Substituting (\ref{Weyl}) 
in (\ref{pi1}) and integrating over ${\bf q}$ and ${\bf q}'$ we obtain
\begin{eqnarray}
\Pi_{loc}({\bf x}',{\bf x};\omega) =
\int \frac{d{\bf p}_0 d{\bf p}_1}{h^{2d}} \delta 
({\bf p}- \frac{ {\bf p}_0+{\bf p}_1}{2})\delta ({\bf p}- {\bf p'}) \nonumber
\\ \times
\frac{ e^{\frac{i}{\hbar}({\bf p}_0-{\bf p}_1) \cdot
({\bf r'}-{\bf r})}}{[\epsilon + \hbar\omega + i0 -{\cal H}({\bf x}_0)]
[\epsilon - i0 -{\cal H}({\bf x}_1)]},~~~~~~~ \nonumber 
\end{eqnarray}
where ${\bf x}_0=(({\bf r}+{\bf r'})/2,{\bf p}_0)$, and 
${\bf x}_1=(({\bf r}+{\bf r'})/2,{\bf p}_1)$. Next, we change variables to
${\bf k'}= ({\bf p}_0+ {\bf p}_1)/2$, ${\bf k}= {\bf p}_0-{\bf p}_1$,
and integrate over ${\bf k'}$. The above integral, then, reduces to
\begin{eqnarray}
\int \frac{d{\bf k}}{h^{2d}}
\frac{\delta ({\bf p}- {\bf p'}) e^{\frac{i}{\hbar}{\bf k} \cdot
({\bf r'}-{\bf r})}}{[\epsilon + \hbar\omega + i0 -{\cal H}({\bf x}^+)]
[\epsilon - i0 -{\cal H}({\bf x}^-)]},\nonumber 
\end{eqnarray}
where ${\bf x}^{\pm}=(({\bf r}+{\bf r'})/2,{\bf p}\pm {\bf k}/2)$.
In the semiclassical limit, $\hbar \to 0$, the integral 
over ${\bf k}$ yields a $\delta$-function, thus
\begin{eqnarray}
\Pi_{loc}({\bf x}',{\bf x};\omega) \simeq \frac{ 
\delta( {\bf x'}-{\bf x})/h^d}{(\epsilon \!+\! \hbar \omega\!+\! i0 
\!-\!{\cal H}({\bf x}))
(\epsilon \!-\! i0 \!-\! {\cal H}({\bf x}))}. \label{piloc1}
\end{eqnarray}
To extract the leading order semiclassical approximation of this expression,
we manipulate the denominator as:
\begin{eqnarray}
den &=&\frac{1}{\hbar \omega + i 0}\left[ 
\frac{1}{\epsilon \!-\! i0 \!-\! {\cal H}({\bf x})} - 
\frac{1}{\epsilon \!+\! \hbar \omega\!+\! i0 \!-\!{\cal H}({\bf x})} \right]
\nonumber \\
&\simeq & \frac{1}{\hbar \omega + i 0}\left[ 
\frac{1}{\epsilon \!-\! i0 \!-\! {\cal H}({\bf x})} - 
\frac{1}{\epsilon \!+\! i0 \!-\!{\cal H}({\bf x})} \right]\nonumber \\
&=& \frac{1}{\hbar \omega + i 0} 2 \pi i
\delta (\epsilon -{\cal H}({\bf x})) \nonumber
\end{eqnarray}
Substituting these results in (\ref{piloc1}), we obtain (\ref{localpi}).
\newline
%%%%%%%%%%%%%%%%%%%%%%%%%%%%%%%%%%%%%%%%%%%%%%%%%%%%%%%%%%%%%%%%%%%%%
\appendix{\bf Appendix B: The classical sum rule (\ref{sumrule})}
\newline

In this appendix we derive the sum rule (\ref{sumrule}).
The derivation will be performed by calculating the right hand side of the 
equation and showing it equals to the left hand side. It is convenient 
to introduce a local coordinate system  with time coordinate, $\tau$,
along the trajectory, and ${\bf r}_\perp$ 
perpendicular to the trajectory. The corresponding conjugate momenta are
the Hamiltonian function $H$, and ${\bf p}_\perp$.
In these coordinates the right hand side of formula (\ref{sumrule}) 
takes the form:
\begin{eqnarray}
rhs &=& \int_0^\infty dt \int
\frac{d Hd H'}{\dot{r}\dot{r}'} d{\bf p}_{\perp}
 d{\bf p}_{\perp}' g({\bf x}, {\bf x}', t)
\delta (\epsilon-H')\nonumber
\\ &\times&
 \delta(H- H'_t)\delta (\tau-\tau_t') 
\delta({\bf r}_\perp-{\bf r'}_{\perp t}) 
\delta({\bf p}_\perp- {\bf p}_{\perp t}').\nonumber
\end{eqnarray}
where $\dot{r}'$ and $\dot{r}$ are the absolute values of the velocity at
the initial and final points respectively, while the subscript $t$ denotes the
value of the corresponding coordinate after time $t$. Thus, in particular,
${\bf r'}_{\perp t}$ and ${\bf p'}_{\perp t}$ are functions 
of the initial phase space point ${\bf x}'=({\bf r',p'})$ and  the time $t$.
Since energy is conserved, one has  $H'_t = H'$.
 The factor $1/\dot{r}\dot{r}'$ in the above integral comes from
the Jacobian of the transformation of variables. 
An integration over $H$, $H'$ and ${\bf p}_{\perp}$ gives
\begin{eqnarray}
rhs &=& \int_0^\infty dt \int 
\frac{d{\bf p}_{\perp}' }{\dot{r}\dot{r}'} \delta(\tau-\tau_t')
\delta({\bf r}_\perp-{\bf r}_{\perp t}') \nonumber \\
& \times & g({\bf x}, {\bf x}', t)|_{H=H'=\epsilon, ~{\bf p}_\perp = 
{\bf p}_{\perp t}'}. \nonumber
\end{eqnarray}  
Since the initial and final points of the particle as well as its 
energy are fixed, there is only a discrete set of trajectories which contribute
to the above integral. These are the trajectories for which ${\bf p}_{\perp}'$
takes a value such that ${\bf r}_\perp={\bf r}_{\perp t}'$.
We denote these trajectories by a subscript $\mu$.
Thus, ${\bf p}_{\perp \mu}'$ and ${\bf p}_{\perp \mu}$ are the
initial and the final momenta of the $\mu$-th trajectory, 
and $t_\mu$ is the corresponding period. Then
\begin{eqnarray}
rhs &=& \sum_\mu \int_0^\infty \!\!\!\!\!dt~ \delta(t\!-\!t_\mu)
\!\! \int_{\Gamma_\mu} 
\!\!\frac{d{\bf p}_{\perp}' }{\dot{r}\dot{r}'}
\delta({\bf r}_\perp\!-\!{\bf r}_{\perp t}') g({\bf x}_\mu, {\bf x}_\mu', t),
\nonumber
\end{eqnarray}     
where $\Gamma_\mu$ is an infinitesimal region surrounding  
${\bf p}_{\perp \mu}'$
in the momentum  space, ${\bf x}_\mu=({\bf r},{\bf p}_\mu)$, and
 ${\bf x}_\mu'=({\bf r}',{\bf p}_\mu')$. 
The integral is straightforward, and the result is
\begin{eqnarray}
rhs &=& \sum_\mu g({\bf x}_\mu,{\bf x}_\mu',t_\mu)
 \left[ \frac{1}{\dot{r} \dot{r}' }
\mbox{Det}^{-1} \left( \frac{\partial {\bf r}_{\perp}}
{\partial {\bf p}_\perp'} \right) \right]_\mu, \nonumber
\end{eqnarray}
where all quantities in the square brackets are calculated for the $\mu$-th 
orbit. 

To see that this result is equal to the left hand side of Eq. (\ref{sumrule}) 
we notice that the expression in the square brackets is related to
the amplitude $A_\mu$ of the semiclassical Green function (\ref{osc}) 
according to\cite{Gutzwiller90}
\begin{eqnarray}
|A_{\mu}|^2 = \frac{1}{h^{d-1}\hbar^2}  \left[\frac{1}{\dot{r} \dot{r}' }
\mbox{Det}^{-1} \left( \frac{\partial {\bf r}_{\perp}}
{\partial {\bf p'}_\perp} \right) \right]_\mu. \label{amu}
\end{eqnarray}
Combining the last two equations yields the sum rule (\ref{sumrule}).
\newline
%%%%%%%%%%%%%%%%%%%%%%%%%%%%%%%%%%%%%%%%%%%%%%%%%%%%%%%%%%%%%%%%%%%%%
\appendix{\bf Appendix C: The product formula Eq.~(\ref{Gproduct})}
\newline

In this appendix we prove the semiclassical formula (\ref{Gproduct})
relating a Green function to the product of two Green functions.
We consider the  case of $G^+_{osc}$. The generalization for  
$G^-_{osc}$ is straightforward. We start by evaluating the
integral on the right hand side of (\ref{Gproduct}) using stationary 
phase approximation. Substituting (\ref{osc}) in (\ref{Gproduct}) yields 
\begin{eqnarray}
rhs= \hbar \sum_{\mu, \mu'} \int \! d{\bf r}_\perp~ \dot{r}
A_{\mu} A_{\mu'} e^{\frac{i}{\hbar}[ S_{\mu} ({\bf r}_1, {\bf r};\epsilon)
+S_{\mu'} ({\bf r}, {\bf r}_0;\epsilon)]}, \label{rhs1}
\end{eqnarray}
where $\mu$ denotes the orbits from ${\bf r}_0$ to ${\bf r}$, while 
 $\mu'$ the orbits from ${\bf r}$ to ${\bf r}_1$. The amplitudes $A_\mu$ and
$A_{\mu'}$ satisfy relation (\ref{amu}). 
The stationary phase condition of the above integral is 
\begin{eqnarray}
\frac{\partial}{\partial{\bf r}_\perp}
\left[ S_{\mu'} ({\bf r}_1, {\bf r};\epsilon)+
S_\mu ({\bf r}, {\bf r}_0;\epsilon) \right] =0.  \nonumber
\end{eqnarray}
With the classical relations $\partial 
S_\mu ({\bf r}, {\bf r}_0;\epsilon)/\partial{\bf r}=
{\bf p}_\mu$ and $S_\mu' ({\bf r}_1, {\bf r};\epsilon)/\partial{\bf r}=
-{\bf p}_\mu'$, it
implies that  the momenta of the  
$\mu$-th and $\mu'$-th  trajectories at ${\bf r}$ are equal,
${\bf p}_\mu = {\bf p}_{\mu'}$.
Thus, ${\bf r}$ lies on a classical trajectory which goes from ${\bf r}_0$ 
to ${\bf r}_1$. We denote this trajectory by $\nu$, see Fig.~10.
Notice that the stationary phase condition
applies only for the perpendicular components of the momenta. However
since the energy is fixed and equal for both trajectories, the longitudinal
components of the momenta of both trajectories have to be equal as well.
{\narrowtext
\begin{figure}[h]
\epsfxsize=8.0cm
%\vspace{-0.5 cm}
\epsfbox{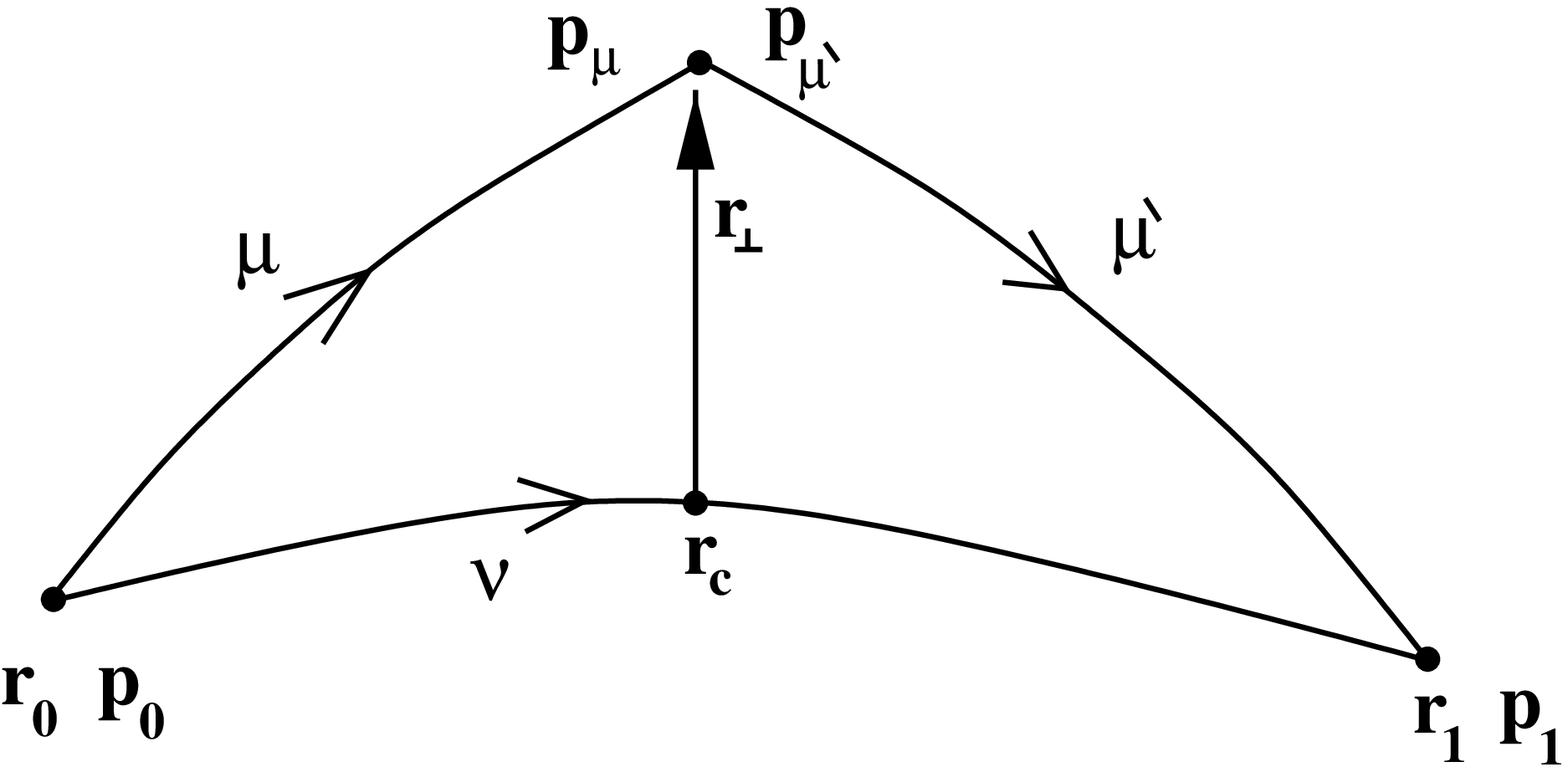}
\vspace{0.5 cm}
\caption{Classical paths joining ${\bf r}_0$ ${\bf r}_1$ and ${\bf r}_c+ 
{\bf r}_\perp$.} 
\end{figure}
}
Denoting by ${\bf r}_c$ a stationary value of {\bf r}, we see that the 
stationary phase is the action of the $\nu$-th orbit from
${\bf r}_0$ to ${\bf r}_1$:
\begin{eqnarray}
S_{\mu'} ({\bf r}_1, {\bf r}_c;\epsilon)+ S_{\mu} 
({\bf r}_c, {\bf r}_0;\epsilon)
= S_\nu({\bf r}_1, {\bf r}_0;\epsilon). \nonumber
\end{eqnarray} 
Thus, the integral (\ref{rhs1}) takes the form  
\begin{eqnarray}
rhs= \sum_\nu I_\nu e^{\frac{i}{\hbar}S_\nu({\bf r}_1, {\bf r}_0;\epsilon)},
\nonumber
\end{eqnarray}
where the pre-exponent factor, $I_\nu$, is  given by
\begin{eqnarray}
I_\nu =\hbar \dot{r} A_\mu A_{\mu'} h^{\frac{d-1}{2}}
\left( \mbox{Det} \left( \frac{\partial {\bf p}_{\perp \mu}-
{\bf p}_{\perp \mu'}}{\partial {\bf r}_\perp} \right) \right)^{-\frac{1}{2}}.
 \label{proda}
\end{eqnarray}
It was calculated by expanding the sum of actions,
$ S_{\mu'} ({\bf r}_1, {\bf r};\epsilon)+ S_{\mu}({\bf r},{\bf r}_0;\epsilon)$,
to second order in ${\bf r}_\perp$ around ${\bf r}_c$, and calculating the
resulting Gaussian integral (\ref{rhs1}).

To see the equivalence between this result and the left hand side of 
(\ref{Gproduct}) it remains to show that $I_\nu=A_\nu$
where $A_{\nu}$ is the amplitude corresponding to the $\nu$-th orbit
from   ${\bf r}_0$ to ${\bf r}_1$. It is 
sufficient to analyze the magnitudes of $I_\nu$ and $A_\nu$, since
the Maslov phases are additive.  Substituting (\ref{amu}), for $A_\mu$ and
$A_{\mu'}$, in (\ref{proda}) we obtain:
\begin{eqnarray}
I_\nu = \frac{1}{\sqrt{h^{d-1} h^2 \dot{r}_0 \dot{r}_1}}
\left\{ \frac{\mbox{Det} \left( \frac{\partial {\bf p}_{\perp\mu'}}
{\partial {\bf r}_{\perp1}} \right)
\mbox{Det} \left( \frac{\partial {\bf p}_{\perp0}}
{\partial {\bf r}_{\perp}} \right) }{
\mbox{Det} \left( \frac{\partial( {\bf p}_{\perp \mu}-
{\bf p}_{\perp \mu'})}{\partial {\bf r}_\perp} \right)} \right\}^{1/2}. \nonumber
\end{eqnarray}
Thus $I_\nu=A_\nu$ if the  term in the curly
brackets equals to $\mbox{Det}(\partial {\bf p}_{\perp 1}/
\partial {\bf r}_{\perp 0}) $.
Following Berry and Mount\cite{Berry72} we first write this term as a 
determinant of the product of the three matrices:
\begin{eqnarray}
\{ ~ \}= \mbox{Det} \left[ \left(  \frac{\partial {\bf p}_{\perp \mu}-
{\bf p}_{\perp \mu'}}{\partial {\bf r}_\perp} \right)^{-1}
\times \frac{\partial {\bf p}_{\perp\mu'}}
{\partial {\bf r}_{\perp1}} \times \frac{\partial {\bf p}_{\perp0}}
{\partial {\bf r}_{\perp}}
 \right]. \label{brakets}
\end{eqnarray}
Now, we use the stationary phase condition 
${\bf p}_{\perp \mu} - {\bf p}_{\perp\mu'}=0$,
which applies at ${\bf r}={\bf r}_c$, and differentiate it with respect to
${\bf r}_{\perp 1}$ keeping ${\bf r}_{\perp 0}$ fixed. Then
\begin{eqnarray}
\frac{\partial {\bf p}_{\perp\mu'}}{\partial {\bf r}_{\perp 1}}
= \frac{\partial ({\bf p}_{\perp \mu} - {\bf p}_{\perp\mu'})}
{\partial {\bf r}_{\perp }} \times \frac{\partial {\bf r}_{\perp }}{
\partial {\bf r}_{\perp 1}}. \nonumber
\end{eqnarray}
Substituting this result in (\ref{brakets}) and using the chain rule of
derivatives we obtain  $\{ ~ \}=\mbox{Det}(\partial{\bf p}_{\perp0}/
\partial {\bf r}_{\perp 1})$ which proves  (\ref{Gproduct}).
\newline
%%%%%%%%%%%%%%%%%%%%%%%%%%%%%%%%%%%%%%%%%%%%%%%%%%%%%%%%%%%%%%%%%%%%%
\appendix{\bf Appendix D: An alternative derivation of (\ref{Zuniversal})}
\newline

In this appendix we derive formula (\ref{Zuniversal}) using random matrix 
theory\cite{Mehta91} (RMT). Our starting point is the Hamiltonian of the form 
${\cal H}_0+ i \gamma/2$, where ${\cal H}_0$ is an $N\times N$ random matrix
of the GOE ensemble, and $\gamma$ is a constant which equals to the typical 
width
of the resonances. With these definitions, $\sigma(\epsilon)= \langle \phi 
| \delta_{\gamma/2}(\epsilon - {\cal H}_0)| \phi \rangle$, 
where 
\begin{eqnarray}
\delta_\gamma (x)= \frac{1}{\pi} \frac{ \gamma}{\gamma^2+ x^2},
\end{eqnarray} 
and  $|\phi \rangle = {\cal D} | g \rangle$ (see Eq.~\ref{hata}).
Denoting by $\varphi_\alpha (j)$ the $j$-th component of eigenfunction
number $\alpha$ of ${\cal H}_0$ we have
\begin{eqnarray}
\sigma(\epsilon)=\eta \sum_\alpha \sum_{jk}\delta_{\gamma/2}
(\epsilon -\epsilon_\alpha)
\varphi_\alpha(j) \varphi_\alpha (k) \phi(j) \phi(k). \label{sigma1}
\end{eqnarray} 
To perform the ensemble averaging we use the property of
RMT that eigenenergies $\epsilon_\alpha$ and the wave functions
$\varphi_\alpha(j)$ are statistically independent. Thus
\begin{eqnarray}
\langle \sigma(\epsilon) \rangle =
\frac{\eta }{N \Delta} \sum_{jk} \phi(j) \phi(k) \delta_{jk}, \nonumber
\end{eqnarray}
where $ 1/\Delta = \langle \sum_\alpha\delta_{\gamma/2}
(\epsilon -\epsilon_\alpha) \rangle$ is the average density of states, 
and $\langle 
\varphi_\alpha(j) \varphi_\alpha (k) \rangle = \delta_{jk}/N$.
Working in units where $\eta \langle \phi| \phi \rangle = N$, we obtain
$\langle \sigma \rangle = 1/\Delta$.

Consider now the connected part of the correlation function
$\langle \sigma(\epsilon) \sigma(\epsilon+ \omega) \rangle$. 
Its calculation requires the average of a
product of two expressions like (\ref{sigma1}).
Here we use the RMT relation:
\begin{eqnarray}
N^2 \langle \varphi_\alpha(j) \varphi_\alpha (k) 
\varphi_\beta(i) \varphi_\beta (l) \rangle =
\delta_{\alpha \beta}[\delta_{ji} \delta_{kl}
+\delta_{jl} \delta_{ki}]+ \delta_{jk} \delta_{il}. \nonumber
\end{eqnarray}
The three terms on the right hand side of the above equation correspond
to the various paring possibilities of orbits as shown in Fig.~2.
In particular, the first and the second terms correspond to Figs.~2a 
and 2b, while the third term  corresponds to Figs.~2c and 2d.
Performing the sum over $i,~j,~k$, and $l$, 
the contribution of the first two terms is 
$I_{ab}(\omega)=2\langle \sum_{\alpha }\delta_{\gamma/2}
(\epsilon -\epsilon_\alpha)\delta_{\gamma/2}
(\epsilon+\omega -\epsilon_\alpha) \rangle$, while
the contribution of the third term is given by 
 the correlator
$I_{cd}(\omega)=\langle \sum_{\alpha \beta}\delta_{\gamma/2}
(\epsilon -\epsilon_\alpha)\delta_{\gamma/2}
(\epsilon+\omega -\epsilon_\beta) \rangle - 1/\Delta^2$.
Thus,
\begin{eqnarray}
I_{ab}(\omega) = \frac{2}{\Delta}\int d\epsilon
\delta_{\gamma/2}(\epsilon)  \delta_{\gamma/2}(\epsilon+\omega) 
= \frac{2 }{\Delta \pi}
\frac{\gamma}{\omega^2+ \gamma^2}, \nonumber
\end{eqnarray}
and
\begin{eqnarray}
I_{cd}(\omega) = \int d\epsilon d\epsilon' 
\delta_{\gamma/2}(\epsilon)  \delta_{\gamma/2}(\epsilon'-\omega) 
R(\epsilon - \epsilon'), \nonumber
\end{eqnarray}
where $R(\epsilon - \epsilon')$ is the two-point function for the density of 
states of the GOE ensemble. Substituting the approximation
$R(\epsilon-\epsilon') \simeq  \Re \int dt t 
e^{i (\epsilon-\epsilon')t}/\pi^2$, 
and performing the integrals over $\epsilon$ and $\epsilon'$ we immediately 
obtain
\begin{eqnarray}
I_{cd}(\omega) = \frac{1}{\pi^2} \Re \int_0^\infty dt~ t e^{-\gamma t + 
i \omega t} =\frac{1}{\pi^2}
\frac{\gamma^2 -\omega^2}{(\gamma^2+\omega^2)^2}. \nonumber
\end{eqnarray}
Collecting these results we arrive at
\begin{eqnarray}
Z^{(0)}(\omega)= \Delta^2 (I_{ab}(\omega)+I_{cd}(\omega)) = ~~~~~~~~~
 \nonumber \\
=\frac{2\Delta}{\pi} 
\frac{\gamma}{\omega^2+ \gamma^2} +
\frac{\Delta^2}{\pi^2} \frac{\gamma^2 -\omega^2}{(\gamma^2+\omega^2)^2}.
\nonumber
\end{eqnarray}
Formula (\ref{Zuniversal}), apart from the prefactor $\xi$, is obtained by
rescaling the variables in the above equation as  
$\Gamma= \gamma/\Delta$ and $\Omega=\omega/\Delta$.

We show, now, that in systems with a separation of time 
scales, $\gamma_1 \gg \gamma_0$, the constant $\xi$ in (\ref{Zuniversal})
is approximately unity. 
The separation of time scales 
implies that for times shorter than $1/\gamma_0$ the dynamics of the 
system may be approximated by that of a close chaotic system
whose Hamiltonian will be denoted by ${\cal H}_0({\bf x})$. 
In this case the first eigenfunctions of the classical propagator
are constant functions corresponding to the ergodic distribution on 
the energy shell, namely $\chi^l({\bf x})= \chi^r({\bf x}) = 
[\int d{\bf x} \delta(\epsilon -{\cal H}_0({\bf x})]^{-1/2}$.
Substituting these eigenfunctions into the expressions 
(\ref{a}) and (\ref{b}) for $a$ and $b$ and evaluating
$\xi=a^2/\langle \langle \rho_\phi \rangle \rangle^2/b$, 
one obtains $\xi=1$.

\end{multicols} 

\begin{references}
\bibitem{comment1} The connected part of the correlator 
of two random variables, say $g$ and $f$, is defined as
$\langle g f \rangle_c=\langle g f \rangle -
\langle g \rangle \langle f \rangle$.
\bibitem{comment2} The Ruelle resonances of the 4-disk system have been
calculated  by Gaspard and Ramirez\cite{Gaspard92} using the cycle expansion.
In this calculation, the ratio of the distance between the disks to 
their radius is 3, while in the experiment\cite{Lu99} it is 4. 
Nevertheless, these theoretical calculations show that indeed 
the subleading Ruelle resonance are complex.
\end{references}
\end{document}